\documentclass[12pt,preprint]{aastex}

\def\arcsecpoint{$''\!.$}
\def\deg{$^{\circ}$}

\shorttitle{Probing the NLR in NGC\,4151}
\shortauthors{ et al.}

\begin{document}

\title{Probing the Ionization Structure of the Narrow Line Region in the Seyfert 1 Galaxy NGC 4151} 

\author{S. B. Kraemer\altaffilmark{1}, H. R. Schmitt\altaffilmark{2},
D. M. Crenshaw\altaffilmark{3}}
\altaffiltext{1}{Institute for Astrophysics and Computational Sciences,
Department of Physics, The Catholic University of America, Washington, DC 20064; and
Astrophysics Science Division, NASA Goddard Space Flight Center, Greenbelt,
MD 20771; kraemer@yancey.gsfc.nasa.gov.} 
\altaffiltext{2}{Remote Sensing Division, Naval Research Laboratory, Washington, DC 20375; and Interferometrics, Inc., Herndon, VA 20171; henrique.schmitt@nrl.navy.mil}
\altaffiltext{3}{Department of Physics and Astronomy, Georgia State University,
Astronomy Offices, Atlanta, GA 30303; crenshaw@chara.gsu.edu}

\begin{abstract}
We present a study of the distribution of [O~III] $\lambda$5007 and
[O~II] $\lambda$3727 emission in the Narrow Line Region (NLR) of the Seyfert~1
galaxy NGC 4151. While the NLR of NGC 4151 exhibits an overall structure
consistent with the unified model of Seyfert galaxies, narrow-band [O~III] and
[O~II] images obtained with the Wide Field and Planetary Camera 2 aboard
the {\it Hubble Space Telescope (HST)} reveal significant emission from outside the
the emission-line bi-cone. The [O~III]/[O~II] ratios are lower in these regions, consistent with a
weaker ionizing flux. We performed a photoionization modeling analysis of the
emission-line gas within a series of annuli, centered on the the central
continuum source, with inner radii from 13 to 90 pc.  The gas is ionized by
radiation that has been attenuated by a relatively highly-ionized absorber (HABS), which completely
covers the central source, and a lower-ionization absorber (LABS), which has a covering
factor ranging from 0 to 1. We found that the [O~III]/[O~II] ratios are well fit
by assuming that, within each segment of an annulus,
some fraction of the NLR gas is completely within the shadow of LABS,
while the rest is irradiated by the continuum filtered
only by HABS.  This suggests that the structure of the NLR is due to filtering
of the ionizing radiation by ionized gas, consistent with disk-wind models. One possible
scenario is that the low-ionization absorbers are dense knots of gas swept
up by a wind.

\end{abstract}

\keywords{galaxies: individual (NGC 4151) -- galaxies: Seyfert --}

\section{Introduction}

Active Galactic Nuclei (AGN) are thought to be powered by accretion of matter onto
super-massive black holes, which reside at the gravitational centers of the host galaxies.
Seyfert galaxies, relatively low luminosity (L$_{bol}$ $\lesssim$ 10$^{45}$ ergs s$^{-1}$), nearby
($z \lesssim 0.1$) AGN, are typically 
grouped into two classes (Khachikian \& Weedman 1974). The spectra of Seyfert 1 galaxies are characterized by
broad (full width half maximum, FHWM, $\gtrsim$ a few 1000 km s$^{-1}$) permitted lines, narrower (FWHM $\lesssim$
1000 km s$^{-1}$) forbidden lines, and strong, non-stellar optical and UV continua, while
the spectra of Seyfert 2's show narrow permitted and forbidden lines and optical and UV 
continua dominated by the host galaxy. Spectral polarimetry of Seyfert 2s (e.g. Antonucci \&
Miller 1985) revealed the presence of broad permitted lines and non-stellar continua
in polarized light. This discovery led to the  unified model for Seyfert
galaxies (Antonucci 1993), which posits that the difference between the two 
results from the orientation of the active region with the respect
to the observer's line-of-sight. The broad emission line gas and continuum source is 
surrounded by a large column of dusty gas, which is along our line-of-sight to Seyfert 2s
and hence obscures our view of the central active region in those galaxies. 

The region in which the broad emission lines form, the so-called broad line region (BLR), is characterized
by gas with hydrogen number densities $n_{H} \geq 10^8 {\rm cm}^{-3}$, and has typically sizes of  
$<$ a few tens of light-days (e.g., Peterson et al. 2004). The narrow line region (NLR), in which the 
forbidden lines arise, is characterized by lower density gas and can extend several hundreds of parsecs.
Although there has been debate in the past about the role of shocks (e.g., Dopita \& Sutherland 1995)
and radio jet/cloud interactions (e.g. Capetti et al. 1997), based on optical/UV studies
(e.g., Kraemer \& Crenshaw 2000a, 2000b; Kraemer et al. 2000, hereafter K2000), X-ray observations
(Sako et al. 2000; Kinkhabwala et al. 2002), and energy considerations (Schmitt et al. 2002), it is
now well-established that the NLR gas is photoionized by the UV-Xray continuum
radiation emitted by the central source.
Narrow-band [O~III] $\lambda$5007 images of Seyfert galaxies show extended NLRs with a bi-conical
structure, with the apex of the bi-cone near the central continuum source. This was first discovered in ground-based
images of the nearest, most extended sources (Pogge 1987) and later confirmed in {\it Hubble Space Telescope (HST)}
images of the inner few hundred parsecs of the NLR (e.eg., Capetti et al. 1996;
Schmitt \& Kinney 1996; Schmitt et al. 2003). The simplest
explanation for such a morphology is that the ionizing radiation, isotropically emitted by the 
black hole/accretion disk, is collimated by the same dusty gas that obscures our view of 
the central source in Seyfert 2s. If so, the morphology and ionization
structure of the NLR is directly related to the distribution and optical thickness of
the obscuring gas. For example, if the circumnuclear material were in the form of a dense 
molecular torus (e.g. Krolik \& Begelman 1988), the bi-cone should have a sharp edge;
NLR gas would be exposed to unfiltered ionizing radiation or completely shielded by the torus. 
However, if the torus has an extended atmosphere, as suggested by some X-ray studies (Feldmeier et al.
1999), or the circumnuclear gas were in the form of a dusty, disk-driven wind 
(Konigl \& Kartje 1994), there would be a region of the NLR that was exposed to filtered
ionizing radiation. As a result, the ionization state of the NLR gas would gradually decrease
with distance relative to the bi-cone axis. 

The inner face of the putative torus is $\sim$ 1 pc from the central source, depending on the
dust sublimation radius (Barvainis 1987), while 
a disk-wind would form at a smaller radial distance and neither can be resolved.
Some insight has already been obtained via the study of line-of-sight absorption
in Seyfert 1s (see Crenshaw, Kraemer, \& George 2003), suggesting that the UV and X-ray absorbers have high global
covering factors and their column densities increase with increasing polar angle with respect to the 
accretion disk axis. One cannot constrain the distribution of the circumnuclear gas in individual 
AGN via absorption-line studies. However, we can use the large-scale structure of the NLR to probe the physical conditions in
the circumnuclear gas. 

Although NGC 4151 ($cz = 990~ {\rm km}~ {\rm s}^{-1}$), is sometimes referred to as the ``prototypical'' Seyfert 1 galaxy,  
it exhibits an unusually extended
NLR (e.g., Evans et al. 1993) compared to other Type 1s. This NLR morphology is consistent with a
line-of-sight towards the active nucleus of NGC 4151, lying just outside the bi-cone edge (e.g. Evans et al. 1993). 
Kinematic studies (Hutchings et al. 1998; Kaiser et al. 2000), using slitless spectra obtained with
the Space Telescope Imaging Spectrograph (STIS) aboard the {\it HST}, found that the
emission-lines southwest of the nuclear point source are blue-shifted, while those on the northeast side are redshifted.
Assuming the geometry suggested by Evans et al., this is consistent with mass-outflow from the AGN. Crenshaw et al.
(2000) and Das et al. (2005) were able to model the outflow by assuming that the emission-line gas is distributed
within a hollow bi-cone.

NGC 4151 was the first Seyfert galaxy to discovered to have intrinsic absorption. Oke \& Sargent (1968) 
found non-stellar He~I $\lambda$3889 and Anderson \& Kraft (1969) detected H$\beta$ and H$\gamma$ self-absorption.
These lines were blue-shifted with respect to the host galaxy, indicating mass outflow. Observations in the UV 
obtained with the {\it IUE} (Boksenberg et al. 1978) and far-UV using the Hopkins UltraViolet Telesocope (Kriss et al., 1992, 1995)
revealed absorption lines from a wide range
of ionization states, including fine-structure and metastable lines.  Early X-ray spectra
revealed the presence of a large column of absorbing gas (Barr et al. 1977; Holt et al.
1980), which subsequent observations revealed to be ionized (Yaqoob, Warwick,
\& Pounds 1989; George et al. 1998). One possible reason for the
complexity of the intrinsic absorption in NGC 4151 is the fact that we are viewing the black hole/accretion
disk system at a relatively high inclination of $\sim$ 45\deg~with respect to the accretion disk (Evans et al. 1993; Das et al. 2005), and 
are likely detecting material close to the densest part of the outflow.
The first high-resolution ($\sim$ 15 km s$^{-1}$) UV spectra
obtained by Weymann et al. (1997) using the Goddard High Resolution Spectrograph (GHRS) aboard {\it HST} revealed
six major kinematic components of C~IV and Mg~II absorption, which were stable over the
period 1992--1996.  We obtained STIS 
spectra in 1999 July (Kraemer et al. 2001) that revealed that the kinematic components detected
by Weymann et al. were still present, although the strongest kinematic components, D and E using the nomenclature of
Weymann et al.,  were broad and we were unable to separate them.
Also, we found complex  
line-of sight covering factors
in the broad D+E component. From our photoionization modeling analysis (see, also, Kraemer et al. 2006) we argued that
the complex, low ionization absorption arose in dense gas, approximately 0.1 pc from the central source.

We obtained STIS spectra using the 0\arcsecpoint1$\times$52$''$ slit, covering the brightest sections of the
NLR, at position angles (PAs) 221\deg and 70\deg, centered on the optical nucleus (Nelson et al. 2000).
Using photoionization models,  
we examined the physical condition in the emission-line gas along the slit (K2000).
The NLR gas can be characterized as consisting of two components: a radiation-bounded component in which lines
from low ionization species, such as [O~II] $\lambda$3727 and [N~II] $\lambda\lambda$6548, 6584, are formed, and
a more highly-ionized, matter-bounded component in which lines such as [Ne~V] $\lambda\lambda$3346, 3426
arise. Both components contribute to the Balmer lines and the strongest of the forbidden lines, [O~III] $\lambda$5007.
The densities of both components decrease with increasing radial distance.
However, the models required the ionizing radiation
to have been absorbed by gas closer to the AGN, which is optically thick at the
He~II Lyman limit, as suggested by Alexander et al. (1999). In fact, the intervening absorbers
were similar to the intrinsic absorbers detected in UV and X-ray spectra of Seyfert
galaxies (George et al. 1998; Crenshaw et al. 1999), although the absorbing gas detected in NGC 4151 is more 
optically thick than that observed in most other Seyfert 1s. 

In Crenshaw \& Kraemer (2007), we argued that the physical
conditions in the emission-line gas in the unresolved knot near the optical nucleus of NGC 4151 are
similar to those of the main sub-component of the absorber D$+$E. In order the match the observed
line luminosities, this component must have a high global covering factor. Emission-line gas in the shadow
of this component would be irradiated by a heavily absorbed continuum.

Expanding on our previous studies, we can begin to test the suggestion that the morphology and ionization
state of the NLR gas is directly related to variations in the attenuation of the ionizing
radiation by gas close to the central source. If the emission-line gas within the bi-cone is
the most highly ionized, it follows that it is exposed to the least-attenuated continuum. 
Furthermore, if the attenuation increases with increasing distance
from the bi-cone axis, we should observe a corresponding drop in ionization. Outside the bi-cone, e.g. along our line-of-sight 
to the central source, the radiation is heavily attenuated, which means that any emission-line gas
must be in a low-ionization
state. Although long-slit spectra are only available for PAs 221\deg and 70\deg, there are
archival {\it HST} Wide Field and Planetary Camera 2 (WFPC2) narrow band [O~III] $\lambda$5007 and
[O~II] $\lambda$3727 images. The [O~III] $\lambda$5007/[O~II] $\lambda$3727 ratio is indicative 
of the ionization state of the gas (e.g. Ferland \& Netzer 1983). Previous photoionization modeling of spatially 
resolved spectra of NGC 4151, and other Seyfert galaxies, 
have been limited to narrow regions within
the emission-line bi-cone, using {\it HST} spectra (e.g. K2000), or the extended NLR (e.g., Schulz \& Komossa 1993).
The present analysis is the first attempt to constrain the physical conditions of the 
entire inner $\sim$ 100 pc of the NLR. Using these images and photoionization
models based on our long-slit study, we will demonstrate that
the variations in [O~III]/[O~II] can be used to map the attenuation of the ionization radiation
as a function of PA.

\section{Observations and Reductions}

The observations used in this paper were obtained with the WFPC2 PC camera
on {it HST}, which has a pixel size of 0.0445\arcsec, corresponding to $\sim$2.8~pc
at the distance of this galaxy, 13.3~Mpc. These observations were done as part
of the project GO-5124 (P.I. Ford), on 1995-Jan-22. A log of the observations
is presented in Table~1, where we can see that each filter had at least one
long exposure, as well as a short one with a duration of 10~s, which is used
to correct for saturated pixels at the nucleus.
 
We retrieved the images from the HST archive,
calibrating them with the best bias, dark and flat-field reference files
available. The remaining reduction steps were done with IRAF using standard
techniques. All images were aligned, and, for those bands with 2 long
exposures ([O~II], [O~III] and [O~II] continuum) we combined the images to
improve their final S/N and eliminate cosmic rays and CCD cosmetic defects. In
the case of the [O~III] continuum image, observed with filter F547M, we had
only one long exposure, so we first eliminated as many cosmic rays as possible
automatically and later eliminated by hand the missed ones. The short 10~s
exposures were also inspected for cosmic ray hits close to the nucleus and
corrected when needed. All images were background subtracted and the 10~s
exposures were scaled to the exposure time of the combined images of their
corresponding filters. These scaled images were used to correct the saturated
nuclear pixels and columns affected by charge bleed in the longer exposures.

The saturation corrected images were flux calibrated using the image header
keywords and the off-band images were used to subtract the host galaxy
continuum contribution from the on-band images. This was done by scaling
the flux of the continuum image based on the width of the two filters.
We checked whether this process over or under subtracted the host galaxy,
by inspecting the outer regions of the pure line images. This flux mismatch
can be caused by the continuum slope between the two filters, especially in
the blue part of the spectrum. Whenever necessary further corrections were
applied, to reduce these residuals. We estimate that the uncertainty in the
continuum subtraction is of the order of 5\%.

The accuracy of the flux calibration of our images was verified by comparing
their fluxes to those obtained with STIS longslit spectroscopy (Nelson et
al. 2000). The continuum subtracted [O~II] and [O~III] image fluxes were
measured along position angle (PA)$=221^{\circ}$, inside areas corresponding to the same
ones used by Nelson et al. (2000). We discarded regions closer than 0.2\arcsec
from the nucleus because of residuals from the continuum subtraction. We find
a good agreement between the two sets of measurements in the case of the [O~III]
fluxes, with the image showing a flux 7\% higher than the spectra (within the
uncertainties of the 2 measurements). In the case of their [O~II] emission we
find that the image fluxes are 63\% higher than the spectra. We attribute this
discrepancy to uncertainties in the image calibration, which can be large
for blue narrow band filters. The [O~II] image was divided by a 1.63 scaling
factor to take this effect into account.

Finally, we determined the noise of the images in regions free from emission,
which was used to determine the uncertainty in the flux measurements. Also,
using the [O~II] image, the one with the highest noise, we created a mask
that blanked all regions with flux below the 3$\sigma$ level. This mask was
applied to both images, to ensure us that we are not comparing regions with
strong emission in one image with noise in the other.

\section{Emission Line Distribution}

The final [O~III] and [O~II] images are presented in Figures~1 and 2,
respectively. The [O~III] image shows the bi-conical structure that was
previously reported by Pogge (1989) and Evans et al. (1993).
The [O~II] image has a structure similar to that of the [O~III], but due to the
lower sensitivity of WFPC2 at short wavelengths we do not see as much extended
emission as in the case of [O~III]. One particular difference between the two
images is the relatively stronger [O~II] emission along directions perpendicular to the
NLR major axis (PA $\sim150^{\circ}$). 

In order to better see the differences between the two emission line
distributions, we used the continuum subtracted images to create an emission
line ratio [O~III]/[O~II] image, shown in Figure~3. 
We did not apply any foreground Galactic extinction correction to
this image, since this region of the sky has a negligible amount of extinction.
Also, we did not try to correct the emission line ratio for internal reddening,
but this is not likely to introduce a large effect on the observed emission
line ratios, since there is very little internal reddening in the NLR of
this galaxy (Nelson et al. 2000; K2000). This figure confirms
the results from the comparison between Figures 1 and 2, where regions along
the cone axis have an [O~III]/[O~II] ratio as much as 10 times higher than 
regions along a perpendicular direction (PA $\sim150^{\circ}$).

Our line-of-sight to the AGN lies outside the bi-cone (Das et al. 2005), therefore,
based on the unified model, the AGN should be obscured by optically-thick, dusty absorption. 
From our UV and X-ray absorption analyses (Kraemer et al. 2005; 2006), we have constraints 
on the physical conditions in the gas along our line-of-sight to the AGN. First, the 
reddening is quite low ($E_{B-V}$ $\approx$ 0.02 mag; Crenshaw \& Kraemer 2005). Also, the most optically thick subcomponent
of D$+$E would allow enough ionizing radiation to
pass through it to ionize material in its shadow. Our line-of-sight lies approximately 12\deg~outside the
outer envelope of the bi-cone (Das et al. 2005). 
This is consistent with Figure 4, which indicates that some ionizing radiation
must escape even in regions outside the extended bi-cone.

\subsection{Flux Measurements}

The goal of this paper is to combine the information obtained from the [O~II]
and [O~III] images with previous spectroscopic results (Nelson et al. 2000;
K2000) and photoionization models, to determine the amount of
filtered and unfiltered ionizing continuum that ionizes the NLR along
different lines of sight. The [O~II] and [O~III] fluxes were measured on the
blanked images, using 15 concentric elliptical annuli with an axial ratio of 0.707
and the major axis oriented along PA $=60^{\circ}$. We have chosen this
geometry so that the ellipse major axes are aligned with the torus axis
and have the same axial ratio as that of the NLR, which is inclined by
45$^{\circ}$ relative to the line of sight (Das et al.  2005).
Using this geometry we can compare regions at similar distances from the
nucleus, differing only in the direction from which the nucleus is seen.
The annuli are 3 pixels wide, starting at a radius of 3 pixels from the nucleus. 
We avoid the inner 3 pixels region because of problems with the continuum
subtraction at the nucleus. Each annulus is divided into 18 sectors with widths
of 20$^{\circ}$. In Figure~4 we show, on the blanked [O~III] image, the distribution
of regions where the measurements were done. Notice the presence of some
sectors where most, or even all the points are blanked. In the following analysis
we eliminate regions where more than $\sim$30\% of the [O~II] or [O~III] pixels 
are blanked.

The flux measurements, emission line ratios and best model fitting
parameters are given in Table~2. The errors
quote in this table correspond to 1~$\sigma$ and were calculated using standard
propagation of errors techniques. Figure~5 shows the distribution of emission
line fluxes and ratios for the inner 8 annuli, as a function of position angle.
Once again we can see the large difference in the excitation state of the
gas in regions along and perpendicular to the NLR axis. Also, the emission-line fluxes
are lower in the latter regions. This is consistent with a drop in the flux of ionizing radiation,
as we will demonstrate in Section 4.

\section{Photoionization Models}

\subsection{Model Input Parameters}

The  photoionization models used for this study were generated using the Beta 5
version of Cloudy (Ferland et al. 1998). We assumed an open, or ``slab'', geometry.
As per convention, the models are parameterized in
terms of $U$, the dimensionless ionization parameter\footnote{
$U = Q/4\pi~r^{2}~c~n_{H}$
where $r$ is the radial distance of the absorber, $n_{H}$ is hydrogen number density, in units of cm$^{-3}$ and $Q = \int_{13.6 eV}^{\infty}(L_{\nu}/h\nu)~d\nu$,
or the number of ionizing photons s$^{-1}$ emitted by a source of luminosity
$L_{\nu}$}, and $N_{H}$, the total hydrogen column density (in units of cm$^{-2}$).
We modeled the intrinsic spectral energy distribution as a broken power law of the form $L_{\nu} \propto 
\nu^{\alpha}$ as follows: $\alpha = -1.0$ for energies $<$ 13.6 eV,
$\alpha = -1.45$ over the range 13.6 eV $\leq$ h$\nu$
$<$ 0.5 keV, and $\alpha = -0.5$ above 0.5  
keV. We included a low energy cut-off at $1.24 \times 10^{-3}$ eV (1 mm) and a high energy cutoff
at 100 keV. The luminosity in ionizing photons was  
$Q = 1.1\times10^{53}$ photons s$^{-1}$. 

Based on our previous photo-ionization analysis (K2000), we assumed
the following elemental abundances, in logarithm, relative to H 
by number: He: $-1.00$, C: $-3.47$, N: $-3.92$, O: $-3.17$, Ne: $-3.96$,
Na; $-5.69$, Mg: $-4.48$,  Al: $-5.53$, 
Si: $-4.51$,  P: $-6.43$, S: $-4.82$,  Ar: $-5.40$, 
Ca: $-5.64$,  Fe: $-4.40$, and Ni: $-5.75$. The heavy element abundances are roughly 1.4 times solar, except for
nitrogen which is twice solar (see Asplund, Grevesse, \& Sauval 2005), scaled in
the manner suggested by Groves et al. (2003). We included cosmic dust in the form of
graphite grains, with 20\% the dust/gas ratio of the Galactic Interstellar Medium (ISM), and
silicate grains, with 50\% the ISM ratio. The inclusion of grains resulted in the following
depletions of elements from gas phase: C, 25\%; O 15\%; Mg, Si, Fe, Ca, Al, and Ni, 50\%. 

Following K2000, we assumed that the emission-line gas within each extraction bin consists
of two separate components, as described in Section 1. We scaled the fluxes of the two components,
based on the predicted H$\beta$ fluxes, so that 80\% of the H$\beta$ emission comes from the 
higher-density/lower-ionization gas and 20\% from the lower-density/higher-ionization gas. This fixes
the ratio of the areas of the illuminated faces of the two components. We assumed the same 
densities as K2000 for the points 13 pc from the AGN and density law, i.e.,  
$n_{H} \propto r^{-1.65}$. We fixed the column densities of the high- and low-density
components at log($N_{H}$) $= 21.0$ and $20.75$, for which they are radiation-bounded and matter-bounded,
respectively, as discussed in K2000.  The full set of model parameters are listed in Table 3.

K2000 determined that the NLR gas was ionized by a continuum filtered by a high-ionization absorber (HABS),
characterized by log($U$)$ = 0$ and log($N_{H}$)$= 22.5$, and a low-ionization
absorber (LABS), with log($U$)$ = -3$ and log($N_{H}$)$= 19.5$, for the SW side of the nucleus,
and log($N_{H}$)$= 20$, for the NE side. Here, we assumed that HABS covers
the full solid angle subtended by the emission-line gas. The parameters for LABS assumed in K2000
were chosen to model the emission-line gas within the brightest parts of the NLR, and we found that
those column densities were too low to sufficiently attenuate the ionization radiation for regions outside
the bi-cone. Therefore, we modified LABS 
by including only one component, with log($N_{H}$)$= 20.5$, which is screened from the central source by HABS,
which drops the ionization parameter slightly (log($U$)$ = -3.07$). The
effects of the absorbers on the ionizing continuum are shown in Figure 6. 

We found that simply increasing the column densities of LABS or HABS produces a sharp cut-off outside
the bi-cone without achieving [O~III]/[O~II] ratios less than unity. The best fit to the
observed [O~III]/[O~II] ratios and fluxes was obtained by holding the column densities of LABS and HABS fixed while
varying the covering factor of LABS.
For each annulus, we generated two sets of models: the high- and low-ionization components
irradiated by a continuum filtered only by HABS, and the two components irradiated
by the continuum filtered by both absorbers. The emission-line fluxes from the two components
were scaled so that the low- and high-density components for the HABS-only models contribute 20\% and 80\%, respectively, of the 
predicted H$\beta$ fluxes\footnote{These are the predicted fluxes assuming that the
line photons escape from the illuminated face of the slab.}. The same scaling
was then used for the LABS models; the scale-factors are listed in Table 3. This method fixes the emitting surface-areas of the 
emission-line gas at each radial distance.
Variations in the covering factor of LABS were
modeled by changing the fractional contribution from the two sets of models, with the sum of the
fractional contributions fixed at unity (assuming values other than unity would require varying the
ratios of the emitting surface-areas).  Note that, since 
there were too few measurements to accurately scale the predicted fluxes, we did not 
did not model the annuli with inner radii greater than 90 pc.

\subsection{Model Results}

The predicted emission-line fluxes for the models, assuming 0\% and 100\%
covering by LABS, are listed in Table 3. Since we have assumed that density falls more
slowly than $r^{-2}$, there is a general trend to lower-ionization with increasing radial
distance that is independent of the effects of absorption. In the case of full covering, there is 
a negligible contribution
from the higher density gas. The drop in emission-line flux is the result of both the absorption
of the ionizing continuum and the fact that only the lower density, hence lower emissivity,
component contributes significantly.

In order to compare the model results with the observations, we varied the covering factor of LABS
to match the observed [O~II]/[O~III] ratio for each 20$^{\rm o}$ segment in PA within the annulus. 
Once the line ratios in the full set of zones at a particular radial distance have been fit, we computed the mean values of the ratios of the 
predicted and measured [O~II] and [O~III] fluxes, which were then used to scale the predicted fluxes.
Note that this scaling includes two factors: the surface area of the emission-line gas, which, multiplied by the
predicted flux yields the line luminosity from the zone, and the dilution factor of the radiation, due to the distance to
NGC 4151. 
 
In order for the model predictions to be physically consistent, the fraction of ionizing radiation
intercepted by the emission-line
gas must be less than unity. To test this, we computed the ratio of the observed [O~III] luminosity to the 
predicted [O~III] flux, prior to scaling, for the brightest segment within each annulus, which yields the
surface area of the illuminated slab. The ratio of this area to the surface area of a sphere
with a radius corresponding to the inner radius of the annulus, multiplied by $20/360$ to account for the
width of the segment, provides a rough estimate of the fraction of ionizing radiation intercepted. For the annuli with inner radii
of 21 pc to 90 pc, the fractions decrease from 0.40 to 0.02. For the annulus with an inner
radius of 13 pc, the fraction is approximately unity. However, this can easily be rectified
if an additional, denser component is included, as discussed in K2000. Alternatively, it is possible that the emission-line
gas within 13 pc is indeed associated with LABS, since it may have a high covering factor (e.g., Crenshaw \& Kraemer 2007).  
In either case, the model predictions are consistent with our assumed ionizing luminosity and the geometric constraints.
Furthermore, this justifies our decision to ignore the effect of screening by the more distant NLR gas.

The model predictions are shown versus the measured values in Figures 7 -- 16. 
In general, the models replicate the variations in the emission line fluxes
as a function of PA, although there are clear mismatches at specific points. In most of these cases,
the fit could be improved by letting the fraction of ionizing radiation, hence the surface areas, of the emission-line gas vary. For example,
the under-predictions of the fluxes at PA $=$ 330\deg -- 350\deg could be evidence that there is simply less
gas, i.e. a smaller emitting surface-area, in those zones. One peculiarity of the models is the 
narrow range in the predicted [O~II] fluxes for radial distances of 13 and 21 pcs. Because of the
relatively high density of the low-ionization component (see Table 3), collisional de-excitation of
the upper levels, $^{2}D_{3/2}$ and $^{2}D_{5/2}$, of O$^{+}$ is important. For the [O~III] line, the upper level 
($^{1}D_{2}$) has a critical
density of 7.0$\times$10$^{5}$ cm$^{-3}$ (Osterbrock 1989) and, therefore, does not show this effect. It is possible
that some of the [O~II] at these radial points arises in more distant, lower-density gas that appears
co-located due to projection effects. Since our line-of-sight is close to the edge of the nominal
bi-cone (Das et al. 2005), gas a large radial distances can appear to be close to the nucleus. However, in spite
of possible problems with the predictions for [O~II], the variation in the line ratios in the inner zones
is mostly driven by changes in the [O~III] flux (see Figure 5), which are well-matched by the models
(see Figure 7a and 8a).
The problem is not evident in the models for the more
distant radial zones, for which the density of this component is below the critical density of the
the $^{2}D_{3/2}$ level of O$^{+}$, n$_{e} = 1.6 \times 10^{4}$ cm$^{-3}$.  

\section{Constraints on the Structure and Distribution of the Absorbers}

As discussed in Section 3, the narrow-band images (Figures 1 and 2) reveal clear evidence for 
[O~III] and [O~II] emission outside the outer boundary of the emission-line bi-cone.
 Furthermore, there does not seem to be
a sharp edge to the bi-cone, as would be expected if the ionizing radiation were
collimated by an optically thick medium (either a torus or a thick wind).
Based on the success of this relatively simple model in replicating the observed
emission-line fluxes, we suggest that the best explanation for the variation in
ionization state across the NLR is attenuation by an optically thick absorber, whose
covering factor increases with distance from the bi-cone axis. 

The variation in the attenuation of the ionizing radiation is illustrated in Figure 17.
Within the bi-cone, i.e. for PAs 30$^{\rm o}$  -- 100$^{\rm o}$ and 200$^{\rm o}$ -- 300$^{\rm o}$, the models
require a relatively low covering factor for LABS, while the covering factor exceeds 0.8 over the
remaining range in PA. This is, of course, suggested by the bi-conical morphology of the NLR in NGC 4151.
More importantly, the covering factor in the more attenuated regions is less than
unity; some ionizing radiation must reach the gas outside the bi-cone. Note, also, that, in order
to fit the [O~II] fluxes, the covering factor
of LABS within the bi-cone is fairly large for the annuli with radii $<$ 40 pc. As mentioned in
the previous section, this is possibly due to projection effects. A similar superposition of components 
was suggested for the inner NLR in NGC 1068 (Kraemer \& Crenshaw 2000a). 
As shown in Figure 17, the covering fraction of LABS does not increase with radial distance. This suggests
that the attenuation of the ionizing radiation occurs close to the AGN, rather than by the NLR gas itself, which is
consistent with our modeling assumptions. Furthermore, the covering fraction of the radiation-bounded NLR gas is 
much less than unity at each radial point (K2000), which makes self-screening by the NLR gas unlikely.

In Figure 18, we show the mean covering factors of LABS, averaged over the set of radial points. The standard 
deviations of the covering factor of 
LABS within the bi-cone are large, which could result from variations in the covering
of the central source on timescales of tens of years. This is not surprising since optically thick, low ionization absorbers 
have been observed to move in and out of our line-of-sight In NGC 4151 over short timescales (e.g. Kraemer et al. 2006).
On the other hand, outside the bi-cone, e.g. PA 100$^{\rm o}$ to 200$^{\rm o}$, the standard deviations are generally lower, which
suggests that the emission-line gas in these regions has experienced little variation in the
covering factor of the intervening absorbers, perhaps due to the larger columns of absorbing gas closer to the
plane of the accretion disk/torus. 
However, there appears to be a relatively bright, high-ionization region between 38 pc and 56 pc, outside the bi-cone near
PA 310\deg (also see Figures 4 and 5), which indicates that there has been some variation in the absorption in these
regions.

However, do our model components and results make sense geometrically? If we assume 
that LABS is $\sim$0.1 pc from the 
nucleus, as indicated in Kraemer et al. (2006), then over a PA interval of 
20\deg, they would have a cumulative lateral size of $\sim$10$^{17}$ cm in 
order to cover $\sim$70\% of the continuum source. At a distance of 30 pc, 
for example, they would cast a shadow of lateral size $\sim$3 $\times$ 
10$^{19}$ cm. Assuming the NLR emission-line clouds are spherical, the 
lateral sizes of the low and high-density clouds are substantially smaller at 
this distance, on the order of 10$^{18}$ and 10$^{17}$cm, respectively (see Table 3). Thus, 
the majority of emission-line clouds would either be completely in or out of 
the shadow cast by the absorbers, as we have modeled them.

Assuming a radial distance of 0.1 pc for both HABS and LABS, their densities and
predicted electron temperatures are, respectively:  10$^{6.75}$ cm$^{-3}$, 8.5 $\times 10^{4}$K;
10$^{9.75}$ cm$^{-3}$, 1.5 $\times 10^{4}$ K. The predicted gas pressures at the ionized
face of each model are 1.52 $\times 10^{-5}$ dynes cm${-2}$ and 2.49 $\times$ 10$^{-2}$ dynes
cm$^{-2}$, hence the components are not in pressure equilibrium, which was also true of the line-of-sight absorbers modeled
by Kraemer et al. (2005). While the column density and ionization parameter we assumed for
HABS is similar to D$+$Ea (Kraemer et al. 2005; 2006), none of UV absorbers were as low in
ionization as LABS. However, during the 1999 July STIS observation of NGC 4151 (Kraemer et al. 2001),
there was strong absorption from metastable Fe~II which was absent during the 2002 May observations (Kraemer et al. 2006),
although the continuum flux was similar during both epochs. Kraemer et al. (2006) suggested this was due to
motion of gas out of out line-of-sight between the two observations. Interestingly, the column densities for
Fe~II, S~II and Ni~II for LABS are similar to those observed in 1999 July (see Table 4). Based on our 
current set of models we suggest that the global covering factor for HABS is near unity, which is consistent
with our previous identification of component D$+$Ea as the source of the unresolved emission line knot 
located near the continuum source (Crenshaw \& Kraemer 2007). LABS may be associated with the low ionization
gas detected in 1999 July, which must also have a large global covering factor to account for the 
variations in the [O~III]/[O~II] ratios. Furthermore, the low-ionization absorbers must have small lateral
sizes and significant transverse velocities (see Kraemer et al. 2006). One possibility is that these are dense
knots entrained by an outflowing wind. In this case, the wind would also have large transverse velocities,
suggesting an origin in the outer parts of the accretion disk (e.g., Kraemer et al. 2005).

\section{Summary} 
 
We have examined the ionization structure of the NLR in NGC 4151 using
narrow-band [O~III] and [O~II] images obtained with {\it HST}/WFPC2.
We measured the line fluxes for a series of annuli centered on the
central continuum source, each divided into extraction bins 
spanning 20\deg~ in PA. We generated photoionization models of the NLR gas for 
annuli with inner radii from 13 to 90 pc. From  our analysis
of the images and photoionization models, we have found the following
regarding the ionization of the NLR and the nature of the ionizing
continuum.

1. Although the overall structure of the NLR is roughly bi-conical, in agreement
with the unified model, there is [O~III] and [O~II] emission from 
regions outside the bi-cone. The [O~III]/[O~II] ratios are lower in these regions,
indicating that the gas is less ionized. Furthermore, the fluxes of both lines
are weaker. Taken together, this indicates that less ionizing radiation reaches these
regions. Based on our previous studies of the intrinsic UV and X-ray absorption
in NGC 4151, our line-of-sight to the AGN is also outside the emission-line bi-cone. However,
our models predict that some ionizing radiation will pass though
the absorbers, which is consistent with the imaging results.

2. We based our characterization of the NLR gas on the results from the analysis of 
STIS longslit spectra (K2000). We assumed that for each extraction bin within an annulus there are
two components of emission-line gas: a higher-density, radiation bounded component and
a lower-density, matter-bounded component. The densities of both components decrease with radial
distance as $r^{-1.65}$. The central source is completely covered by an absorber 
of log$(U) = 0.$ and log$(N_{H}) = 22.5$. As per K2000, we also included a lower-ionization
absorber (LABS) of log$(U) = -3.$ and log$(N_{H}) = 20.5$. We assumed that within any segment of an
annulus, some fraction of the emission-line gas lies in the shadow of the low-ionization
absorber. We varied this fraction by adding the predicted emission-line fluxes for
models generated with and without the effect of LABS until we matched the observed line ratios.
Then, we generated a single factor per bin to scale the predicted fluxes to the observed fluxes. The
scale factor includes the fraction of ionizing radiation intercepted by the emission-line gas,
which is fixed at each radius, and the $1/$distance$^{2}$ dilution
of the line luminosities. The model predictions fit the observations quite accurately, which
confirms that the illumination pattern of the NLR is the result of varying attenuation of the
ionizing continuum and that the low-ionization absorbers have small lateral sizes.

3. The predicted ionic column densities of S~II, Fe~II, and Ni~II from LABS are in rough agreement with those
measured in STIS echelle spectra obtained in 1999 July. The absence of these low ionization species in
subsequent STIS observations, combined with the constraints on the lateral size of LABS, suggests that
this absorption arises in small, dense knots entrained in more highly ionized gas, which has a significant
transverse velocity, as expected if it formed as part of a disk-wind.

To summarize, the ionization structure of the NLR is consistent with attenuation of the ionizing radiation
by clumpy, ionized gas close to the AGN. Although, in principle, the attenuation could occur in the extended atmosphere of
a circumnuclear torus, the similarity between our model absorbers and those observed in our line-of-sight
towards the AGN suggests that the attenuation is by a disk-wind. However, given the complex intrinsic
absorption and small amount of reddening present in NGC 4151, it may be a unique object. In order
to reach more general conclusions about the nature of the circumnuclear gas in AGN, it will
be necessary to examine the detailed ionization structure of the NLRs in other nearby Seyferts.

\acknowledgments

This research made use of the NASA/IPAC Extragalactic Database (NED), which is
operated by the Jet Propulsion Laboratory, Caltech, under contract with NASA.
The observations used in this paper were obtained with the NASA/ESA Hubble
Space Telescope at the Space Telescope Science Institute, which is operated
by the Association of Universities for Research in Astronomy, Inc., under
NASA contract NAS5-26555. Basic research at the US Naval Research Laboratory
is supported by the Office of Naval Research. We thank Gary Ferland for his
continued development and maintenance of ClOUDY.

\clearpage

\begin{deluxetable}{lrll}
\tabletypesize{\scriptsize}
\tablecaption{Observing Log\label{tb-1}}
\tablewidth{0pt}
\tablehead{
\colhead{Dataset Name}&\colhead{Exposure Time}&\colhead{Filter}&\colhead{Comment}\\
\colhead{}&\colhead{(s)}&\colhead{}&\colhead{}}
\startdata
U2I50101T &  260 & F502N & [OIII]$\lambda$5007\\
U2I50102T &  600 & F502N & [OIII]$\lambda$5007\\
U2I50103T &   10 & F502N & [OIII]$\lambda$5007\\
U2I50107T &  300 & F547M & [OIII]$\lambda$5007-cont.\\
U2I50108T &   10 & F547M & [OIII]$\lambda$5007-cont.\\
U2I50109T &  500 & F375N & [OII]$\lambda$3727\\
U2I5010AT & 1200 & F375N & [OII]$\lambda$3727\\
U2I5010BT &   10 & F375N & [OII]$\lambda$3727\\
U2I5010CT &  160 & F336W & [OII]$\lambda$3727-cont.\\
U2I5010DT &  700 & F336W & [OII]$\lambda$3727-cont.\\
U2I5010ET &   10 & F336W & [OII]$\lambda$3727-cont.\\
\enddata
\end{deluxetable}

\begin{deluxetable}{rrrrrrrr}
\tabletypesize{\scriptsize}
\tablecaption{Flux Measurements and Model Parameters\label{tb-2}}
\tablewidth{0pt}
\tablehead{
\colhead{Radius (pc)}& \colhead{PA (degrees)} & \colhead{F([OII]$^{a}$)}& \colhead{F([OIII]$^{b}$)}& \colhead{[OIII]/[OII]}
&\colhead{$\sigma$([OII]$^{c}$)}&\colhead{$\sigma$([OIII]$^{c}$)}&\colhead{$\sigma$([OIIII]/[OII])}}      
\startdata
12.8  &  10   &         2.89 &  2.33  &  8.07 &  0.65 &  0.18  &        0.18    \\
21.4  &  10   &         3.23 &  1.56  &  4.83 &  0.92 &  0.26  &        0.13    \\
29.9  &  10   &         3.54 &  1.53  &  4.33 &  1.30 &  0.37  &        0.15    \\
38.4  &  10   &         2.73 &  1.19  &  4.37 &  1.30 &  0.37  &        0.20    \\
47.0  &  10   &         2.47 &  0.71  &  2.91 &  1.52 &  0.44  &        0.18    \\
55.5  &  10   &         1.83 &  0.33  &  1.80 &  1.65 &  0.47  &        0.16    \\
64.1  &  10   &         0.74 &  0.10  &  1.48 &  1.65 &  0.47  &        0.34    \\
72.6  &  10   &         0.80 &  0.10  &  1.33 &  1.95 &  0.56  &        0.33    \\
12.8  &  30   &         7.28 &  5.13  &  7.04 &  0.92 &  0.26  &        0.08    \\
21.4  &  30   &         4.86 &  2.33  &  4.79 &  1.12 &  0.32  &        0.11    \\
29.9  &  30   &         7.97 &  7.57  &  9.50 &  1.45 &  0.42  &        0.17    \\
38.4  &  30   &         5.18 &  4.50  &  8.71 &  1.45 &  0.42  &        0.24    \\
47.0  &  30   &         6.06 &  2.61  &  4.30 &  1.78 &  0.51  &        0.12    \\
55.5  &  30   &         6.88 &  2.60  &  3.79 &  1.89 &  0.54  &        0.10    \\
64.1  &  30   &         5.79 &  2.10  &  3.62 &  1.95 &  0.56  &        0.12    \\
72.6  &  30   &         3.57 &  1.59  &  4.48 &  2.15 &  0.62  &        0.27    \\
81.2  &  30   &         0.95 &  0.22  &  2.33 &  2.25 &  0.65  &        0.55    \\
12.8  &  50   &         11.66&  4.30  &  3.69 &  1.02 &  0.29  &        0.03    \\
21.4  &  50   &         5.49 &  2.61  &  4.76 &  1.21 &  0.35  &        0.10    \\
29.9  &  50   &         5.38 &  5.12  &  9.53 &  1.45 &  0.42  &        0.25    \\
38.4  &  50   &         4.01 &  3.57  &  8.91 &  1.72 &  0.49  &        0.38    \\
47.0  &  50   &         6.41 &  3.34  &  5.21 &  1.89 &  0.54  &        0.15    \\
55.5  &  50   &         11.78&  5.98  &  5.08 &  2.00 &  0.57  &        0.08    \\
64.1  &  50   &         14.09&  6.14  &  4.35 &  2.20 &  0.63  &        0.06    \\
72.6  &  50   &         9.16 &  4.07  &  4.44 &  2.30 &  0.66  &        0.11    \\
81.2  &  50   &         4.87 &  1.47  &  3.03 &  2.43 &  0.70  &        0.15    \\
12.8  &  70   &         5.63 &  1.98  &  3.52 &  0.92 &  0.26  &        0.05    \\
21.4  &  70   &         5.26 &  3.21  &  6.11 &  1.21 &  0.35  &        0.14    \\
29.9  &  70   &         6.58 &  6.55  &  9.96 &  1.52 &  0.44  &        0.23    \\
38.4  &  70   &         6.34 &  6.39  &  10.08&  1.78 &  0.51  &        0.28    \\
47.0  &  70   &         8.92 &  6.45  &  7.23 &  1.89 &  0.54  &        0.15    \\
55.5  &  70   &         6.80 &  5.18  &  7.61 &  1.95 &  0.56  &        0.21    \\
64.1  &  70   &         10.68&  5.61  &  5.26 &  2.15 &  0.62  &        0.10    \\
72.6  &  70   &         14.41&  6.65  &  4.61 &  2.34 &  0.67  &        0.07    \\
81.2  &  70   &         8.70 &  2.79  &  3.21 &  2.47 &  0.71  &        0.09    \\
89.7  &  70   &         5.18 &  1.64  &  3.18 &  2.64 &  0.76  &        0.16    \\
98.3  &  70   &         1.91 &  0.36  &  1.91 &  2.68 &  0.77  &        0.27    \\
12.8  &  90   &         4.46 &  1.30  &  2.91 &  0.65 &  0.18  &        0.04    \\
21.4  &  90   &         6.19 &  1.76  &  2.85 &  1.30 &  0.37  &        0.06    \\
29.9  &  90   &         5.75 &  2.89  &  5.03 &  1.30 &  0.37  &        0.11    \\
38.4  &  90   &         5.11 &  3.11  &  6.09 &  1.52 &  0.44  &        0.18    \\
47.0  &  90   &         4.39 &  1.66  &  3.78 &  1.72 &  0.49  &        0.14   \\
55.5  &  90   &         4.32 &  1.95  &  4.52 &  1.84 &  0.53  &        0.19    \\
64.1  &  90   &         4.73 &  2.52  &  5.34 &  2.05 &  0.59  &        0.23    \\
72.6  &  90   &         4.68 &  2.28  &  4.88 &  2.10 &  0.60  &        0.22    \\
81.2  &  90   &         3.81 &  1.13  &  2.98 &  2.30 &  0.66  &        0.18    \\
89.7  &  90   &         4.32 &  1.75  &  4.07 &  2.30 &  0.66  &        0.21    \\
98.3  &  90   &         4.39 &  0.88  &  2.01 &  2.43 &  0.70  &        0.11    \\
12.8  &  110  &         7.73 &  2.28  &  2.96 &  0.92 &  0.26  &        0.03    \\
21.4  &  110  &         3.16 &  0.69  &  2.18 &  0.92 &  0.26  &        0.06    \\
29.9  &  110  &         2.31 &  0.57  &  2.47 &  1.21 &  0.35  &        0.13    \\
38.4  &  110  &         2.05 &  0.45  &  2.21 &  1.38 &  0.39  &        0.15    \\
47.0  &  110  &         1.20 &  0.18  &  1.56 &  1.59 &  0.46  &        0.21    \\
55.5  &  110  &         0.60 &  0.08  &  1.36 &  1.59 &  0.46  &        0.36    \\
12.8  &  130  &         8.12 &  1.86  &  2.30 &  0.79 &  0.23  &        0.02    \\
21.4  &  130  &         3.75 &  0.91  &  2.44 &  0.92 &  0.26  &        0.06    \\
29.9  &  130  &         3.15 &  0.54  &  1.71 &  1.12 &  0.32  &        0.06    \\
38.4  &  130  &         2.26 &  0.27  &  1.21 &  1.12 &  0.32  &        0.06    \\
47.0  &  130  &         1.85 &  0.27  &  1.48 &  1.52 &  0.44  &        0.12    \\
55.5  &  130  &         0.85 &  0.09  &  1.14 &  1.59 &  0.46  &        0.22    \\
64.1  &  130  &         0.79 &  0.05  &  0.70 &  1.52 &  0.44  &        0.14    \\
12.8  &  150  &         4.05 &  0.63  &  1.57 &  0.65 &  0.18  &        0.02    \\
21.4  &  150  &         2.64 &  0.39  &  1.49 &  0.79 &  0.23  &        0.04    \\
29.9  &  150  &         2.84 &  0.43  &  1.54 &  1.12 &  0.32  &        0.06    \\
38.4  &  150  &         2.10 &  0.34  &  1.65 &  1.30 &  0.37  &        0.10    \\
47.0  &  150  &         1.48 &  0.26  &  1.75 &  1.30 &  0.37  &        0.15    \\
64.1  &  150  &         0.75 &  0.07  &  0.96 &  1.65 &  0.47  &        0.22    \\
72.6  &  150  &         0.60 &  0.03  &  0.51 &  1.78 &  0.51  &        0.17    \\
81.2  &  150  &         1.32 &  0.05  &  0.39 &  1.72 &  0.49  &        0.06    \\
12.8  &  170  &         3.76 &  0.99  &  2.66 &  0.65 &  0.18  &        0.04    \\
21.4  &  170  &         4.58 &  0.67  &  1.48 &  0.92 &  0.26  &        0.03    \\
29.9  &  170  &         3.27 &  0.60  &  1.86 &  1.12 &  0.32  &        0.06    \\
38.4  &  170  &         3.00 &  0.45  &  1.52 &  1.21 &  0.35  &        0.06    \\
47.0  &  170  &         2.52 &  0.38  &  1.52 &  1.38 &  0.39  &        0.08    \\
55.5  &  170  &         1.41 &  0.19  &  1.37 &  1.52 &  0.44  &        0.15    \\
64.1  &  170  &         1.23 &  0.18  &  1.47 &  1.65 &  0.47  &        0.20    \\
12.8  &  190  &         4.78 &  2.15  &  4.50 &  0.79 &  0.23  &        0.07    \\
21.4  &  190  &         5.63 &  2.40  &  4.27 &  1.12 &  0.32  &        0.08    \\
29.9  &  190  &         3.55 &  1.47  &  4.15 &  1.21 &  0.35  &        0.14    \\
38.4  &  190  &         3.87 &  1.64  &  4.24 &  1.30 &  0.37  &        0.14    \\
47.0  &  190  &         4.05 &  1.07  &  2.64 &  1.65 &  0.47  &        0.10    \\
55.5  &  190  &         3.54 &  0.84  &  2.39 &  1.65 &  0.47  &        0.11    \\
64.1  &  190  &         3.65 &  0.65  &  1.78 &  1.78 &  0.51  &        0.08    \\
72.6  &  190  &         1.38 &  0.19  &  1.39 &  1.95 &  0.56  &        0.20    \\
81.2  &  190  &         1.23 &  0.19  &  1.60 &  1.89 &  0.54  &        0.25    \\
12.8  &  210  &         5.79 &  2.59  &  4.47 &  0.79 &  0.23  &        0.06    \\
21.4  &  210  &         4.23 &  3.36  &  7.94 &  1.12 &  0.32  &        0.21    \\
29.9  &  210  &         5.14 &  4.48  &  8.71 &  1.45 &  0.42  &        0.24    \\
38.4  &  210  &         5.51 &  3.63  &  6.59 &  1.45 &  0.42  &        0.17    \\
47.0  &  210  &         7.50 &  3.14  &  4.19 &  1.78 &  0.51  &        0.10    \\
55.5  &  210  &         6.60 &  3.71  &  5.62 &  1.84 &  0.53  &        0.15    \\
64.1  &  210  &         6.34 &  4.12  &  6.49 &  1.95 &  0.56  &        0.20    \\
72.6  &  210  &         6.82 &  2.47  &  3.63 &  2.15 &  0.62  &        0.11    \\
81.2  &  210  &         3.81 &  1.17  &  3.08 &  2.25 &  0.65  &        0.18    \\
89.7  &  210  &         4.35 &  1.08  &  2.48 &  2.39 &  0.69  &        0.13    \\
98.3  &  210  &         3.22 &  1.33  &  4.15 &  2.43 &  0.70  &        0.31    \\
106.8 &  210  &         5.98 &  2.08  &  3.48 &  2.52 &  0.72  &        0.14    \\
115.3 &  210  &         5.52 &  1.69  &  3.07 &  2.79 &  0.80  &        0.15    \\
123.9 &  210  &         3.62 &  0.75  &  2.09 &  2.72 &  0.78  &        0.15    \\
132.4 &  210  &         0.88 &  0.20  &  2.28 &  2.83 &  0.81  &        0.73    \\
12.8  &  230  &         5.26 &  4.41  &  8.39 &  1.02 &  0.29  &        0.16    \\
21.4  &  230  &         4.46 &  4.92  &  11.02&  1.30 &  0.37  &        0.32    \\
29.9  &  230  &         4.76 &  4.00  &  8.41 &  1.45 &  0.42  &        0.25    \\
38.4  &  230  &         11.51&  7.26  &  6.31 &  1.65 &  0.47  &        0.09    \\
47.0  &  230  &         12.46&  5.53  &  4.44 &  1.89 &  0.54  &        0.06    \\
55.5  &  230  &         11.78&  5.55  &  4.71 &  2.00 &  0.57  &        0.08    \\
64.1  &  230  &         12.37&  7.99  &  6.46 &  2.30 &  0.66  &        0.12    \\
72.6  &  230  &         11.16&  6.92  &  6.20 &  2.30 &  0.66  &        0.12    \\
81.2  &  230  &         9.48 &  3.83  &  4.04 &  2.47 &  0.71  &        0.10    \\
89.7  &  230  &         3.51 &  0.95  &  2.70 &  2.56 &  0.74  &        0.19    \\
115.3 &  230  &         3.14 &  1.29  &  4.11 &  3.01 &  0.87  &        0.39    \\
123.9 &  230  &         5.77 &  2.28  &  3.96 &  3.05 &  0.88  &        0.21    \\
132.4 &  230  &         5.93 &  1.88  &  3.17 &  3.08 &  0.89  &        0.16    \\
12.8  &  250  &         5.52 &  5.00  &  9.07 &  1.02 &  0.29  &        0.16    \\
21.4  &  250  &         7.12 &  4.33  &  6.09 &  1.21 &  0.35  &        0.10    \\
29.9  &  250  &         6.37 &  3.77  &  5.92 &  1.52 &  0.44  &        0.14    \\
38.4  &  250  &         4.37 &  3.50  &  8.03 &  1.65 &  0.47  &        0.30    \\
47.0  &  250  &         8.11 &  6.16  &  7.59 &  1.95 &  0.56  &        0.18    \\
55.5  &  250  &         13.37&  7.40  &  5.53 &  2.05 &  0.59  &        0.08    \\
64.1  &  250  &         18.97&  10.32 &  5.44 &  2.10 &  0.60  &        0.06    \\
72.6  &  250  &         15.53&  7.47  &  4.81 &  2.30 &  0.66  &        0.07    \\
81.2  &  250  &         16.07&  9.49  &  5.90 &  2.52 &  0.72  &        0.09    \\
89.7  &  250  &         15.37&  9.52  &  6.19 &  2.56 &  0.74  &        0.10    \\
98.3  &  250  &         13.57&  6.82  &  5.02 &  2.79 &  0.80  &        0.10    \\
106.8 &  250  &         6.14 &  2.02  &  3.29 &  2.79 &  0.80  &        0.15    \\
12.8  &  270  &         3.32 &  3.45  &  10.38&  0.79 &  0.23  &        0.24    \\
21.4  &  270  &         7.86 &  8.94  &  11.37&  1.21 &  0.35  &        0.17    \\
29.9  &  270  &         9.91 &  6.23  &  6.29 &  1.30 &  0.37  &        0.08    \\
38.4  &  270  &         8.58 &  3.59  &  4.19 &  1.59 &  0.46  &        0.07    \\
47.0  &  270  &         5.07 &  2.55  &  5.04 &  1.65 &  0.47  &        0.16    \\
55.5  &  270  &         11.61&  5.46  &  4.70 &  1.84 &  0.53  &        0.07    \\
64.1  &  270  &         10.27&  5.73  &  5.58 &  2.00 &  0.57  &        0.10    \\
72.6  &  270  &         6.26 &  2.83  &  4.51 &  2.15 &  0.62  &        0.15    \\
81.2  &  270  &         6.24 &  2.19  &  3.51 &  2.30 &  0.66  &        0.13    \\
89.7  &  270  &         4.01 &  1.17  &  2.92 &  2.34 &  0.67  &        0.17    \\
98.3  &  270  &         1.47 &  0.22  &  1.52 &  2.43 &  0.70  &        0.25    \\
106.8 &  270  &         1.45 &  0.16  &  1.15 &  2.60 &  0.75  &        0.21    \\
12.8  &  290  &         4.21 &  6.38  &  15.17&  0.79 &  0.23  &        0.28    \\
21.4  &  290  &         3.89 &  2.37  &  6.08 &  1.02 &  0.29  &        0.16    \\
29.9  &  290  &         7.85 &  5.09  &  6.48 &  1.21 &  0.35  &        0.10    \\
38.4  &  290  &         9.44 &  5.64  &  5.97 &  1.38 &  0.39  &        0.08    \\
47.0  &  290  &         6.14 &  3.56  &  5.79 &  1.52 &  0.44  &        0.14    \\
55.5  &  290  &         4.74 &  1.51  &  3.18 &  1.65 &  0.47  &        0.11    \\
64.1  &  290  &         2.60 &  1.21  &  4.65 &  1.84 &  0.53  &        0.32    \\
72.6  &  290  &         0.77 &  0.36  &  4.76 &  1.84 &  0.53  &        1.13    \\
81.2  &  290  &         1.71 &  0.84  &  4.94 &  2.05 &  0.59  &        0.59    \\
12.8  &  310  &         4.19 &  2.76  &  6.59 &  0.65 &  0.18  &        0.10    \\
21.4  &  310  &         1.83 &  1.08  &  5.90 &  0.79 &  0.23  &        0.25    \\
29.9  &  310  &         3.79 &  2.05  &  5.43 &  1.12 &  0.32  &        0.16    \\
38.4  &  310  &         5.70 &  3.56  &  6.25 &  1.21 &  0.35  &        0.13    \\
47.0  &  310  &         6.65 &  5.01  &  7.53 &  1.52 &  0.44  &        0.17    \\
55.5  &  310  &         3.50 &  2.09  &  5.96 &  1.52 &  0.44  &        0.25    \\
64.1  &  310  &         1.60 &  0.51  &  3.21 &  1.59 &  0.46  &        0.32    \\
72.6  &  310  &         0.84 &  0.12  &  1.48 &  1.78 &  0.51  &        0.31    \\
81.2  &  310  &         0.94 &  0.09  &  0.99 &  1.78 &  0.51  &        0.19    \\
12.8  &  330  &         7.76 &  4.61  &  5.94 &  0.79 &  0.23  &        0.06    \\
21.4  &  330  &         1.79 &  1.25  &  7.00 &  0.92 &  0.26  &        0.35    \\
29.9  &  330  &         1.81 &  0.91  &  5.05 &  1.02 &  0.29  &        0.28    \\
38.4  &  330  &         2.42 &  1.04  &  4.29 &  1.30 &  0.37  &        0.23    \\
47.0  &  330  &         1.58 &  0.89  &  5.66 &  1.30 &  0.37  &        0.46    \\
55.5  &  330  &         1.11 &  0.52  &  4.69 &  1.45 &  0.42  &        0.61    \\
64.1  &  330  &         1.56 &  0.25  &  1.63 &  1.59 &  0.46  &        0.16    \\
72.6  &  330  &         0.83 &  0.08  &  1.06 &  1.65 &  0.47  &        0.21    \\
81.2  &  330  &         0.60 &  0.06  &  0.99 &  1.78 &  0.51  &        0.30    \\
12.8  &  350  &         4.10 &  1.83  &  4.45 &  0.65 &  0.18  &        0.07    \\
21.4  &  350  &         4.95 &  1.56  &  3.15 &  1.02 &  0.29  &        0.06    \\
29.9  &  350  &         2.44 &  0.78  &  3.22 &  1.12 &  0.32  &        0.14    \\
38.4  &  350  &         3.35 &  0.64  &  1.93 &  1.30 &  0.37  &        0.07    \\
47.0  &  350  &         1.54 &  0.41  &  2.65 &  1.38 &  0.39  &        0.23    \\
55.5  &  350  &         1.44 &  0.25  &  1.74 &  1.59 &  0.46  &        0.19    \\
64.1  &  350  &         0.89 &  0.10  &  1.19 &  1.59 &  0.46  &        0.21    \\
72.6  &  350  &         1.10 &  0.08  &  0.74 &  1.78 &  0.51  &        0.12    \\
\enddata
\tablenotetext{a}{in units of 10$^{-15}$ ergs cm$^{-2}$ s$^{-1}$.}
\tablenotetext{b}{in units of 10$^{-14}$ ergs cm$^{-2}$ s$^{-1}$.}
\tablenotetext{c}{in units of 10$^{-16}$ ergs cm$^{-2}$ s$^{-1}$.}
\end{deluxetable}

\begin{deluxetable}{llllllll}
\tabletypesize{\scriptsize}
\tablecaption{Model Parameters and Predicted Fluxes$^{a}$\label{tb-3}}
\tablewidth{0pt}
\tablehead{\colhead{Radial Distance (pc)} 
&\colhead{Absorbers$^{b}$} &\colhead{log($n_{H}$)} &\colhead{log(U)$^{c}$} &\colhead{log($N_{H}$)}
&\colhead{$F$([O~II])} 
&\colhead{$F$([O~III])} &\colhead{Scale-Factor$^{d}$}
}
\startdata
13 & HABS & 4.54 & $-$2.35 & 21.0 & 1.01 & 32.0 & 0.35\\
   & HABS & 3.03 & $-$0.84 & 20.75 & & & \\
   & LABS & 4.54 & $-$4.35 & 21.0 & 1.86 & 1.21 $\times 10^{-2}$ & \\
   & LABS & 3.03 & $-$2.83 & 20.75 & & & \\
21 & HABS & 4.20 & $-$2.43 & 21.0 & 0.72 & 11.6 & 0.71 \\
   & HABS & 2.69 & $-$0.92 & 20.75 & & & \\
   & LABS & 4.20 & $-$4.43 & 21.0 & 0.58 & 4.03 $\times 10^{-3}$ & \\
   & LABS & 2.69 & $-$2.92 & 20.75 & & & \\
30 & HABS & 3.94 & $-$2.48 & 21.0 & 0.52 & 5.39 & 1.31 \\
   & HABS & 2.43 & $-$0.97 & 20.75 &  & & \\
   & LABS & 3.94 & $-$4.48 & 21.0 & 0.25 & 1.17 $\times 10^{-3}$ & \\
   & LABS & 2.43 & $-$2.97 & 20.75 & & &  \\
38 & HABS & 3.78 & $-$2.53 & 21.0 & 0.40 & 3.16 & 1.95\\
   & HABS & 2.27 & $-$1.02 & 20.75 & & & \\
   & LABS & 3.78 & $-$4.53 & 21.0 & 0.13 & 6.38 $\times 10^{-4}$ & \\
   & LABS & 2.27 & $-$3.02 & 20.75 & & & \\
47 & HABS & 3.62 & $-$2.55 & 21.0 & 0.31 & 2.02 & 2.82 \\
   & HABS & 2.11 & $-$1.04 & 20.75 & & & \\
   & LABS & 3.62 & $-$4.55 & 21.0 & 8.30 $\times 10^{-2}$ & 4.00 $\times 20^{-4}$ & \\
   & LABS & 2.11 & $-$3.04 & 20.75 & & &  \\
56 & HABS & 3.49 & $-$2.57 & 21.0 & 0.24 & 1.37 & 3.81 \\
   & HABS & 1.98 & $-$1.06 & 20.75 & & & \\
   & LABS & 3.49 & $-$4.57 & 21.0 & 5.40 $\times 10^{-2}$ & 2.67 $\times 10^{-4}$ & \\
   & LABS & 1.98 & $-$3.06 & 20.75 & & & \\
64 & HABS & 3.39 & $-$2.59 & 21.0 & 0.20 & 1.02 & 4.30 \\
   & HABS & 1.88 & $-$1.08 & 20.75 & & & \\
   & LABS & 3.39 & $-$4.59 & 21.0 & 3.97 $\times 10^{-2}$ & 1.96 $\times 10^{-4}$ & \\
   & LABS & 1.88 & $-$3.08 & 20.75 & & & \\
73 & HABS & 3.30 & $-$2.61 & 21.0 & 0.16 & 0.76 & 4.61 \\
   & HABS & 1.79 & $-$1.10 & 20.75 & & & \\
   & LABS & 3.30 & $-$4.61 & 21.0 & 2.84 $\times 10^{-2}$ & 1.42 $\times 10^{-4}$ & \\
   & LABS & 1.79 & $-$3.11 & 20.75 & & & \\
81 & HABS & 3.22 & $-$2.62 & 21.0 & 0.14 & 0.60 & 6.91 \\
   & HABS & 1.71 & $-$1.11 & 20.75 & & & \\
   & LABS & 3.22 & $-$4.62 & 21.0 & 2.23 $\times 10^{-2}$ & 1.13 $\times 10^{-4}$ & \\
   & LABS & 1.71 & $-$3.11 & 20.75 & & & \\
90 & HABS & 3.15 & $-$2.63 & 21.0 & 0.12 & 0.49 & 10.2\\
   & HABS & 1.64 & $-$1.12 & 20.75 &  & & \\
   & LABS & 3.15 & $-$4.63 & 21.0 & 1.77 $\times 10^{-2}$ & 8.99 $\times 10^{-5}$ & \\
   & LABS & 1.64 & $-$3.13 & 20.75 &\\
\enddata
\tablenotetext{a}{Fluxes are in units of ergs cm$^{-2}$ s$^{-1}$. The values listed are the combined fluxes
for the high- and low-density components, emitted at the ionized face of the slabs (see Section 4.1)}
\tablenotetext{b}{The intervening absorbers, HABS and LABS, are described in Section 4.1. In the case where LABS is listed,
the ionizing continuum is filtered through both absorbers.}
\tablenotetext{c}{The ionization parameter, $U$, is computed by Cloudy, including the effect
of the filtered continuum.}
\tablenotetext{d}{In units of 10$^{-14}$ erg cm$^{-2}$ s$^{-1}$. The scale-factors are used
to scale the predicted emitted flux to match the observed flux. The same scale-factor is
used for all points at the same radial distance.}
\end{deluxetable}

\begin{deluxetable}{lll}
\tabletypesize{\scriptsize}
\tablecaption{Comparison of Observed and Predicted Ionic Columns$^{a}$\label{tb-4}}
\tablewidth{0pt}
\tablehead{
\colhead{ion}&\colhead{1999 STIS}&\colhead{LABS}}
\startdata
S~II & 1.0 $\times 10^{3}$ $\pm$ 5.0 $\times 10^{2}$ & 2.5 $\times 10^{2}$ \\
Fe~II $\lambda$1608 & $>$ 1.7 $\times 10^{2}$ & 7.9 $\times 10^{2}$ \\
Ni~II & 85 $\pm$22 & 40\\
\enddata
\tablenotetext{a}{In units of 10$^{13}$ cm${-2}$.}
\end{deluxetable}

\clearpage

\begin{figure}
\plotone{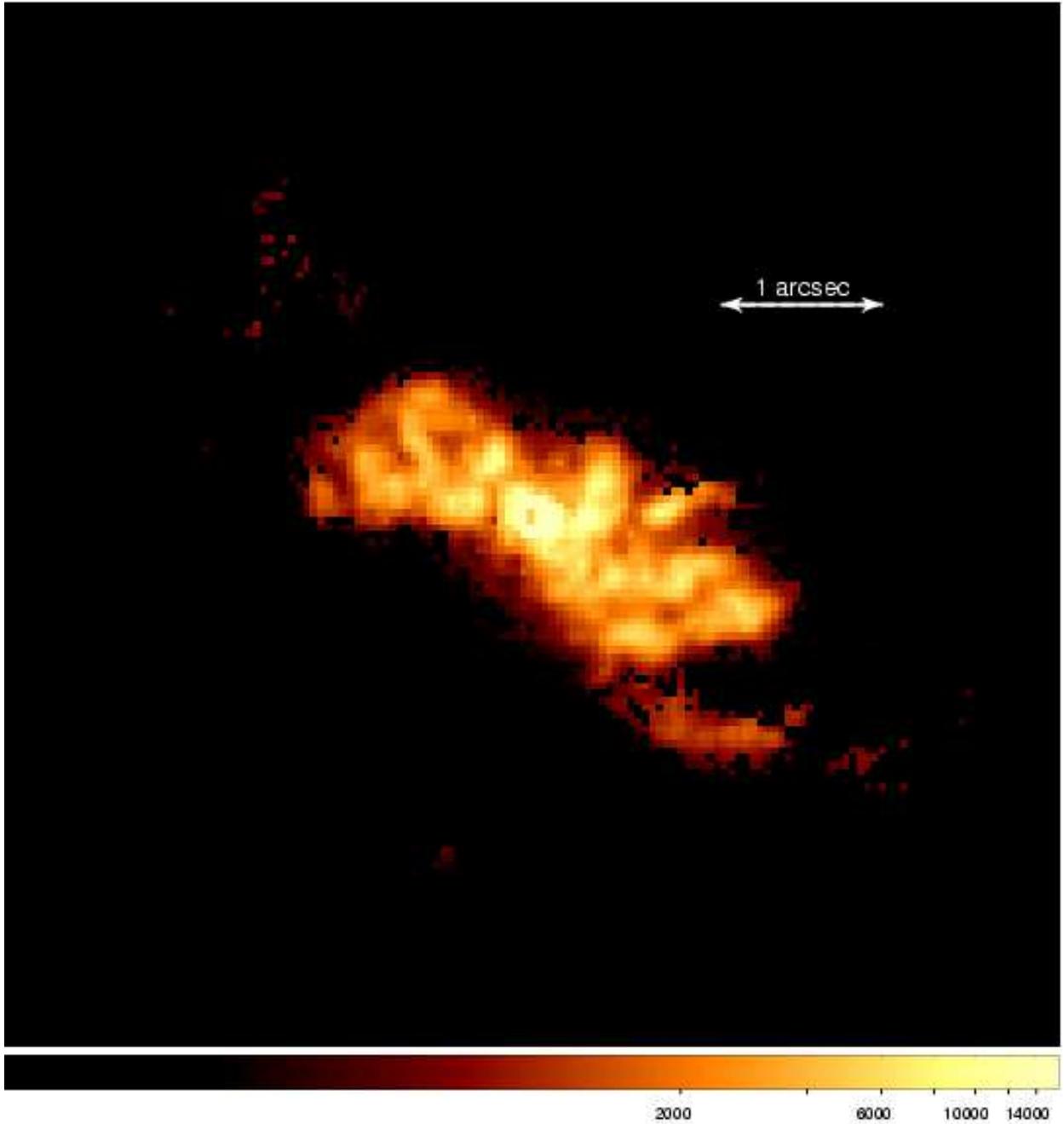}
\caption{Continuum subtracted [O~III] image of NGC\,4151. The orientation is N up,
E to the left, and the horizontal bar indicates the 1\arcsec scale. The image was
blanked to show only those regions with flux higher than 3$\sigma$ above the
background in the [O~II] image. The image is displayed using a logarithmic scale,
which is displayed at the bottom in units of 10$^{-18}$ erg~cm$^{-2}$~s$^{-1}$.}
\end{figure}

\clearpage

\begin{figure}
\plotone{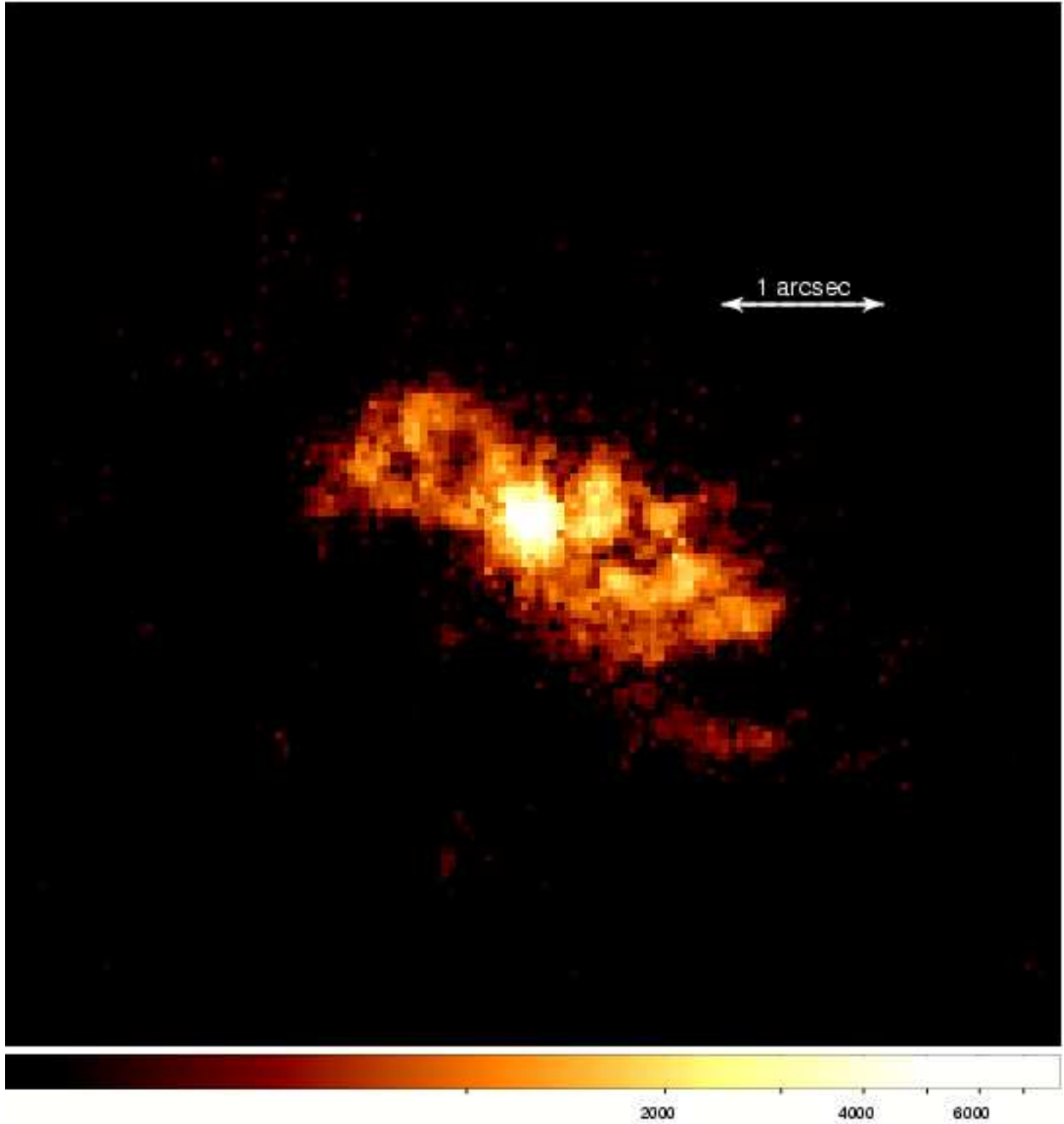}
\caption{Same as Figure 1 for the continuum subtracted [O~II] image.}
\end{figure}

\clearpage

\begin{figure}
\plotone{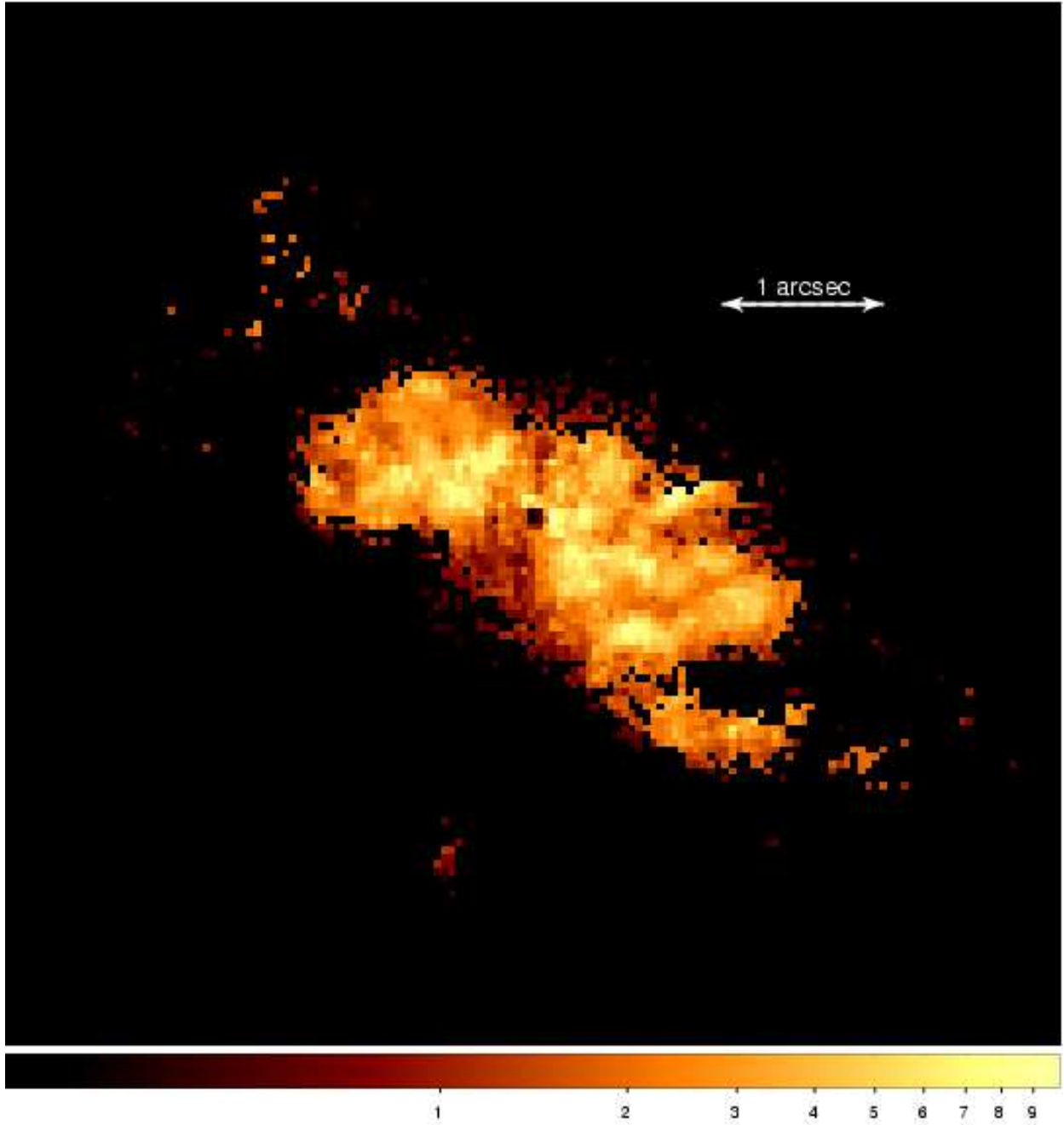}
\caption{Same as Figure 1 for the emission line ratio [O~III]/[O~II].}
\end{figure}

\clearpage

\begin{figure}
\plotone{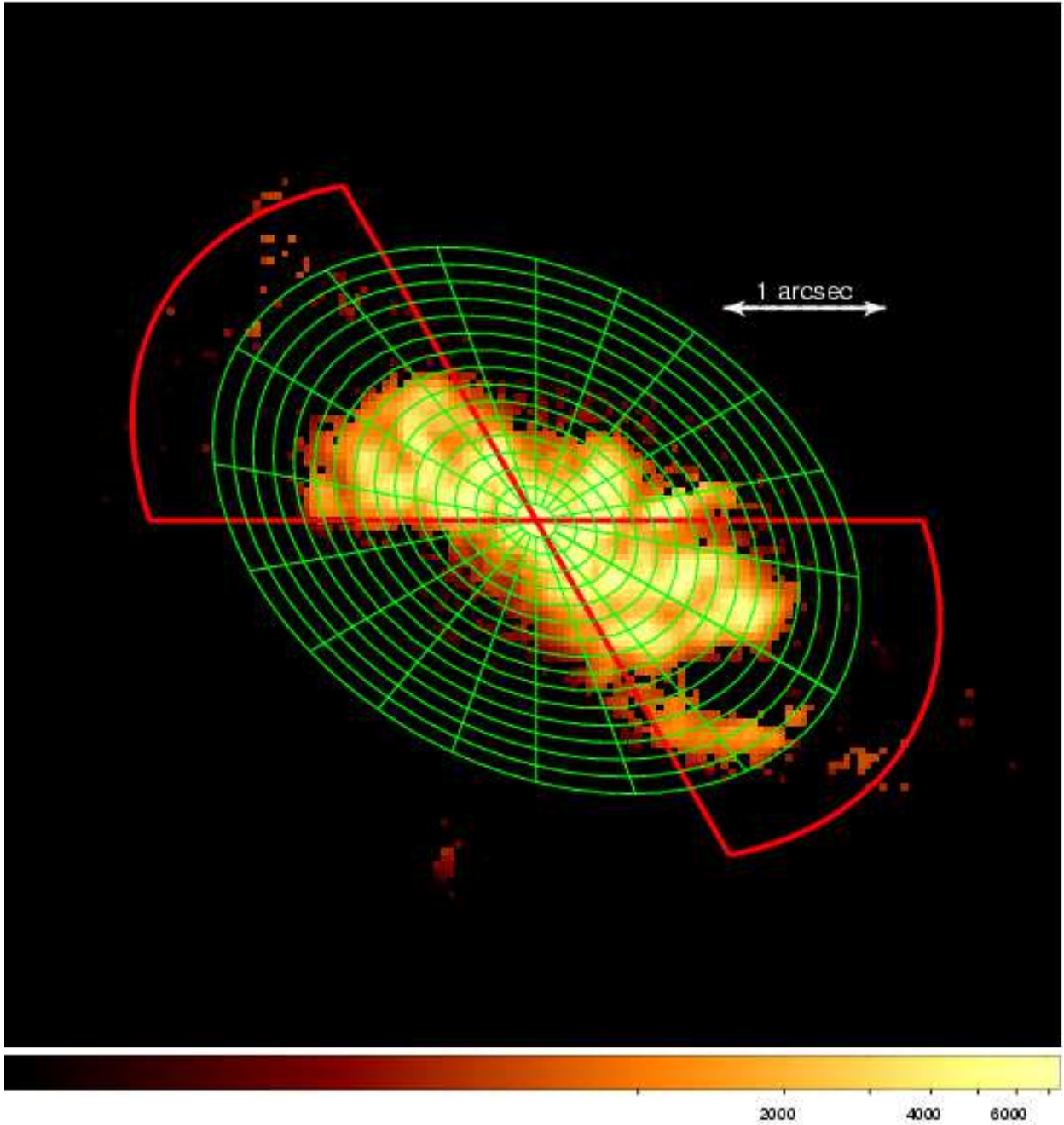}
\caption{Blanked [O~III] image of NGC\,4151 showing in green the distribution of
annuli and sectors used to measure the emission line fluxes of the images
displayed in Figures~1 and 2. We used 15 elliptical annuli with a width of
3~pixels each, starting at a distance of 3 pixels from the nucleus. Each annulus
is divided into 18 sectors of 20$^{\circ}$. The ellipses have an axis ratio 0.707,
corresponding to an inclination of 45$^{\circ}$.
In red we show the NLR bi-cone.}
\end{figure}

\clearpage

\begin{figure}
\plotone{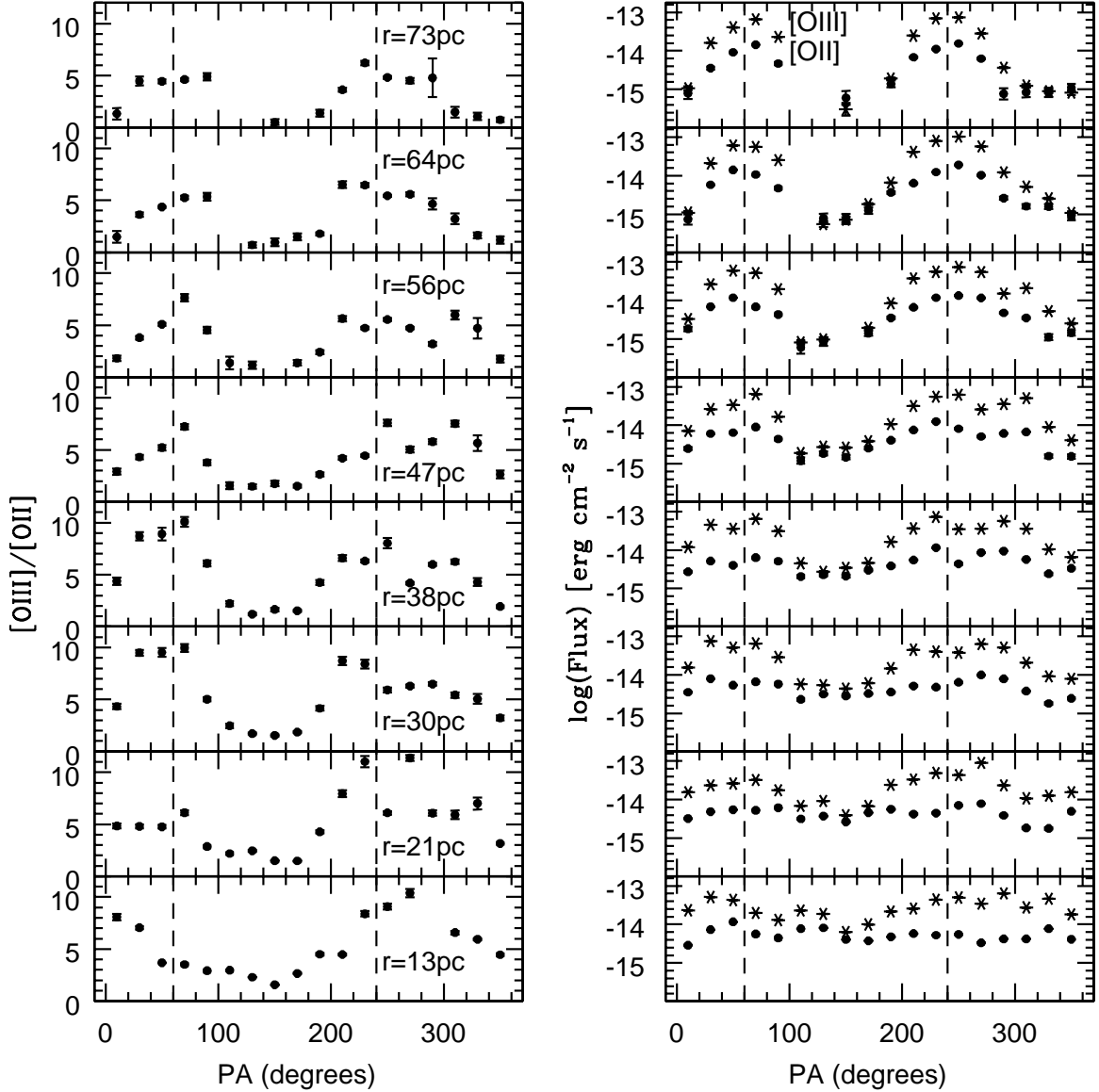}
\caption{The distribution of the [O~III]/[O~II] ratio (left) and the logarithm of
the emission line fluxes (right) as a function of position angle. Each panel
corresponds to different deprojected distances from the nucleus, indicated
in the right portion of each panel. The vertical dashed lines indicate the position
angle of the NLR major axis (torus axis). In the right panel we show the [O~III] and
[O~II] fluxes as stars and dots, respectively. }
\end{figure}

\clearpage

\begin{figure}
\includegraphics[scale=0.75,angle=90]{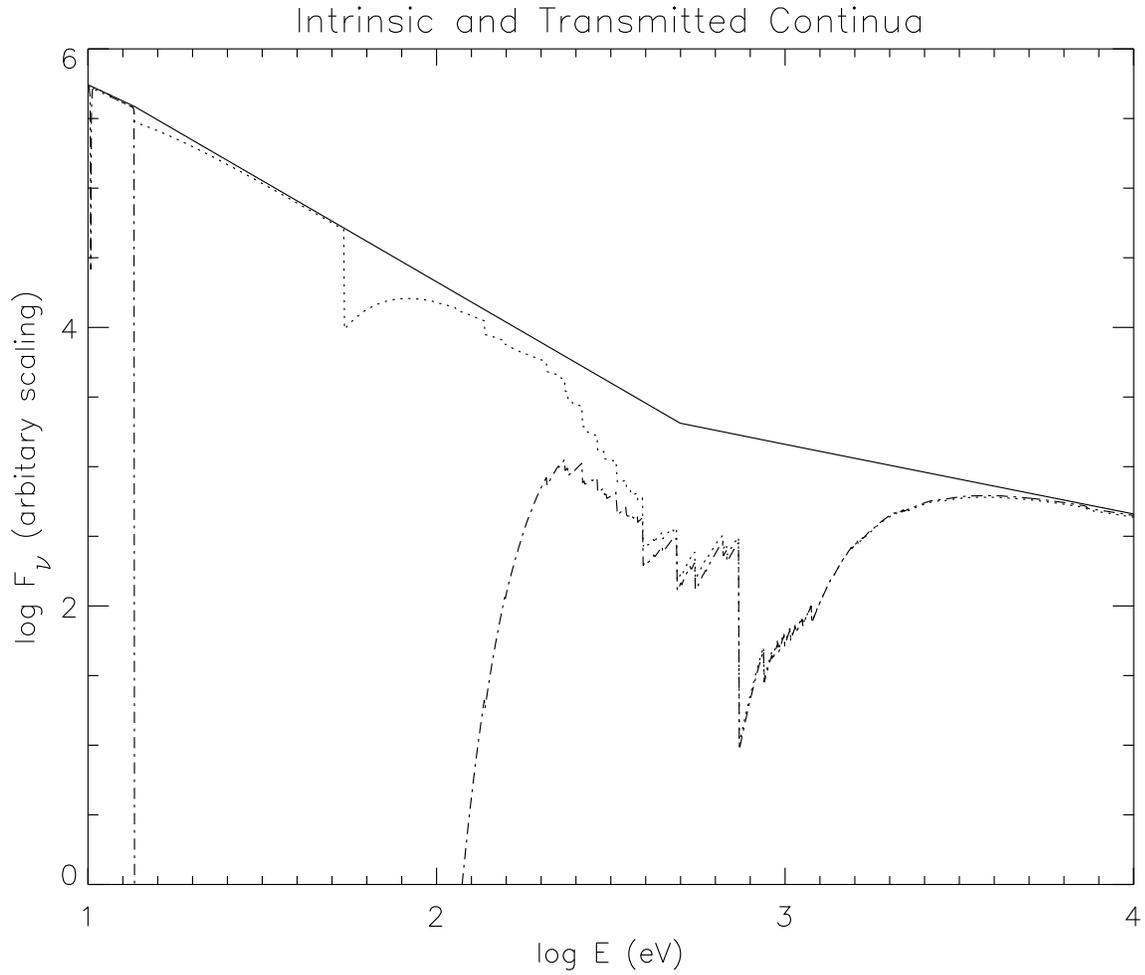}
\caption{The ionizing continua assumed for the photo-ionization modeling. The sold line shows
the intrinsic continuum. The dotted and dash-dotted lines are the continua transmitted through
HABS and LABS, respectively (see text).}
\end{figure}

\clearpage

\begin{figure}
\includegraphics[scale=0.6,angle=90]{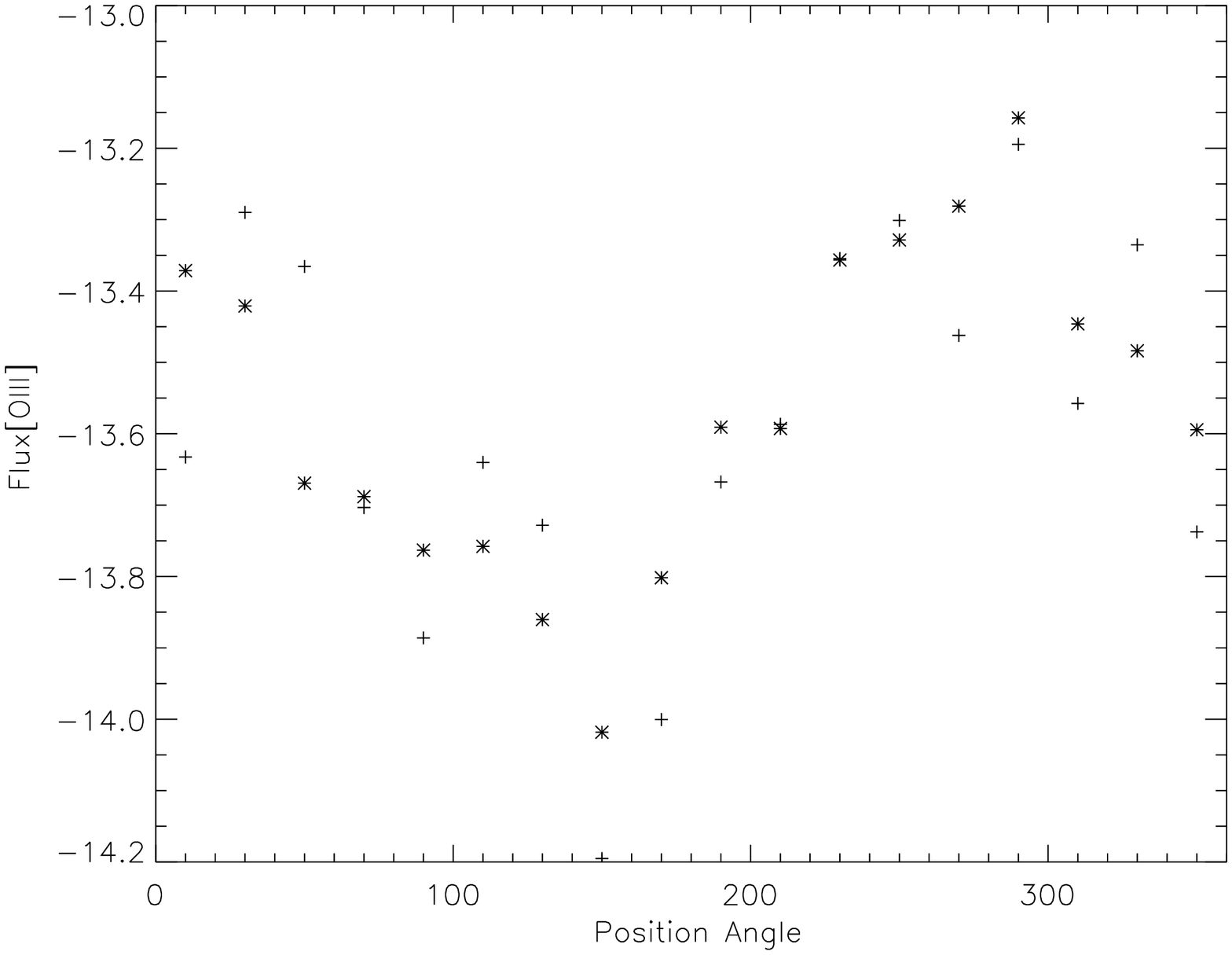}
\includegraphics[scale=0.6,angle=90]{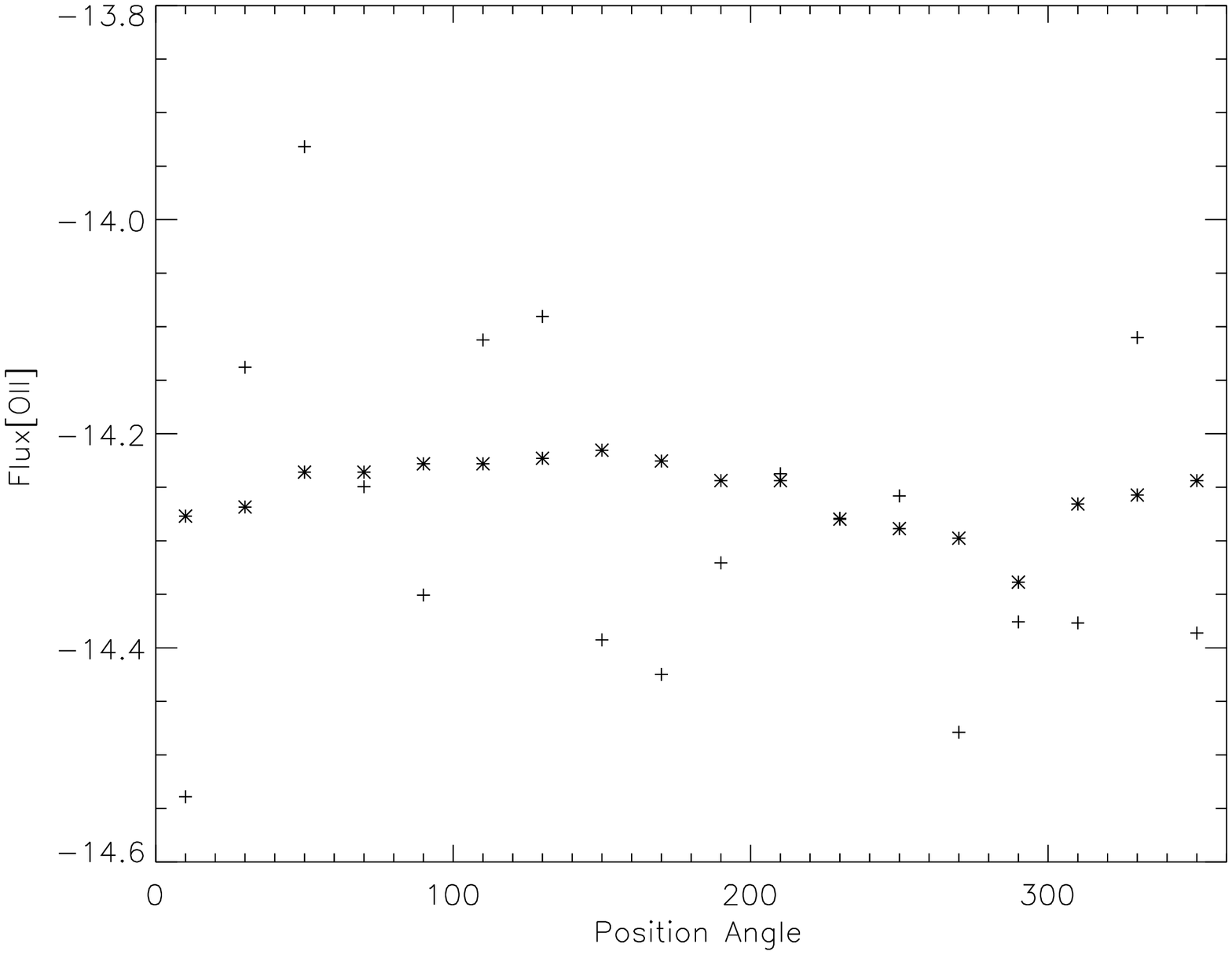}
\caption{{\it Top Panel}: Observed (crosses) and predicted (asterisks) [O~III] fluxes over the
range in position angle (PA) sampled at a radial distance of 13 pc. {\it Bottom Panel}: Observed and predicted [O~II]
fluxes for the same range in PA.}
\end{figure}

\clearpage

\begin{figure}
\includegraphics[scale=0.6,angle=90]{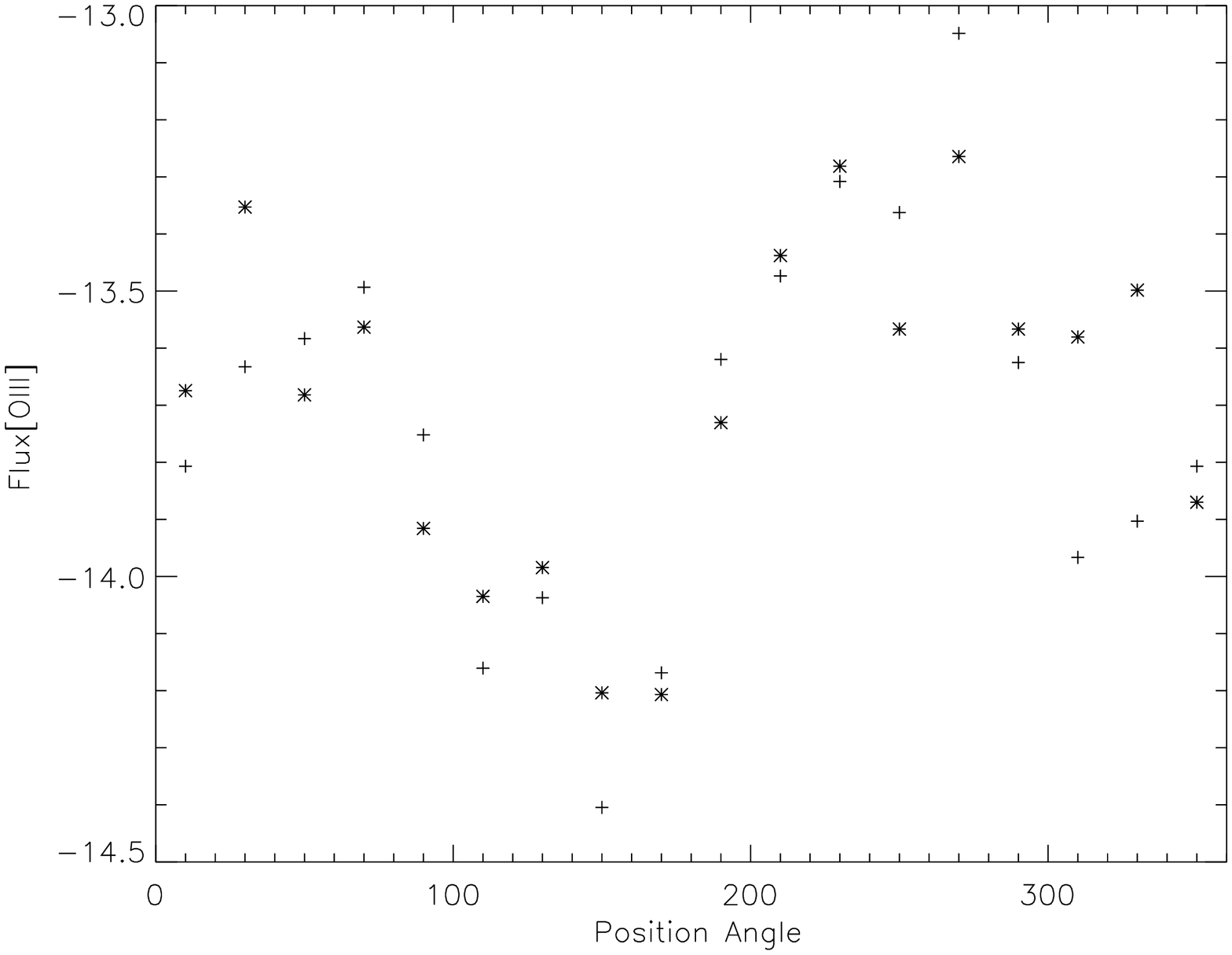}
\includegraphics[scale=0.6,angle=90]{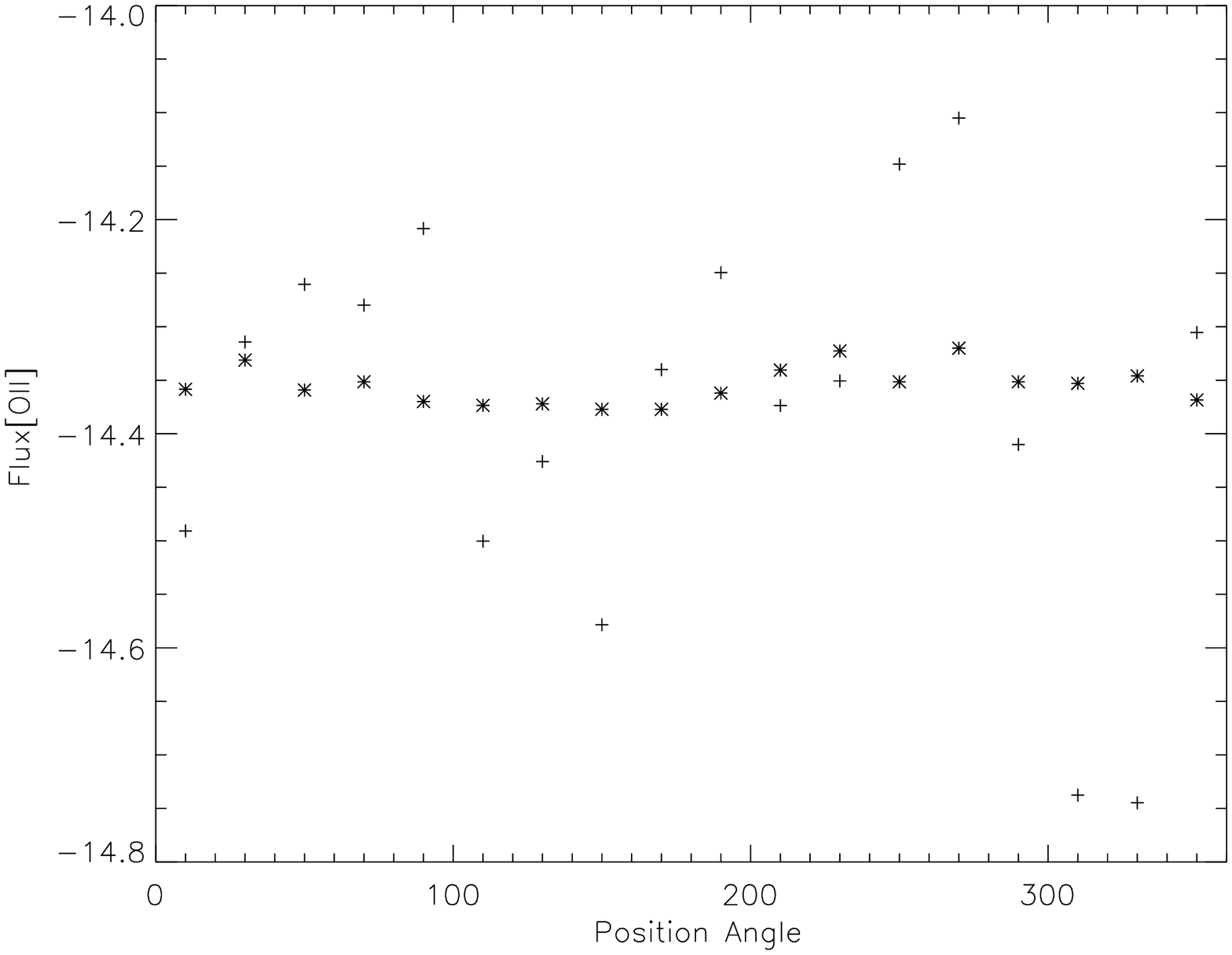}
\caption{{\it Top Panel}: Observed (crosses) and predicted (asterisks) [O~III] fluxes over the
range in position angle (PA) sampled at a radial distance of 21 pc. {\it Bottom Panel}: Observed and predicted [O~II]
fluxes for the same range in PA.}
\end{figure}

\clearpage

\begin{figure}
\includegraphics[scale=0.6,angle=90]{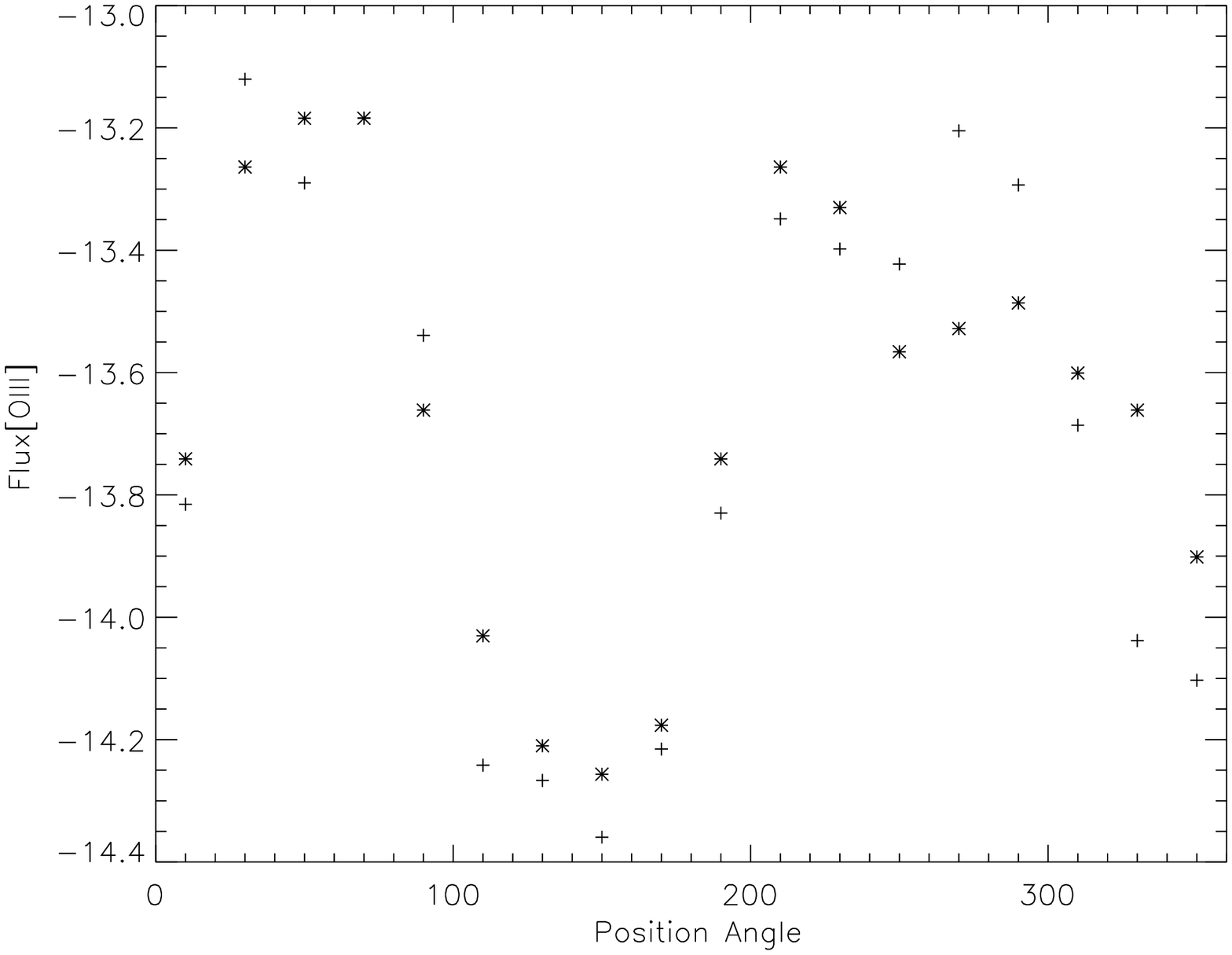}
\includegraphics[scale=0.6,angle=90]{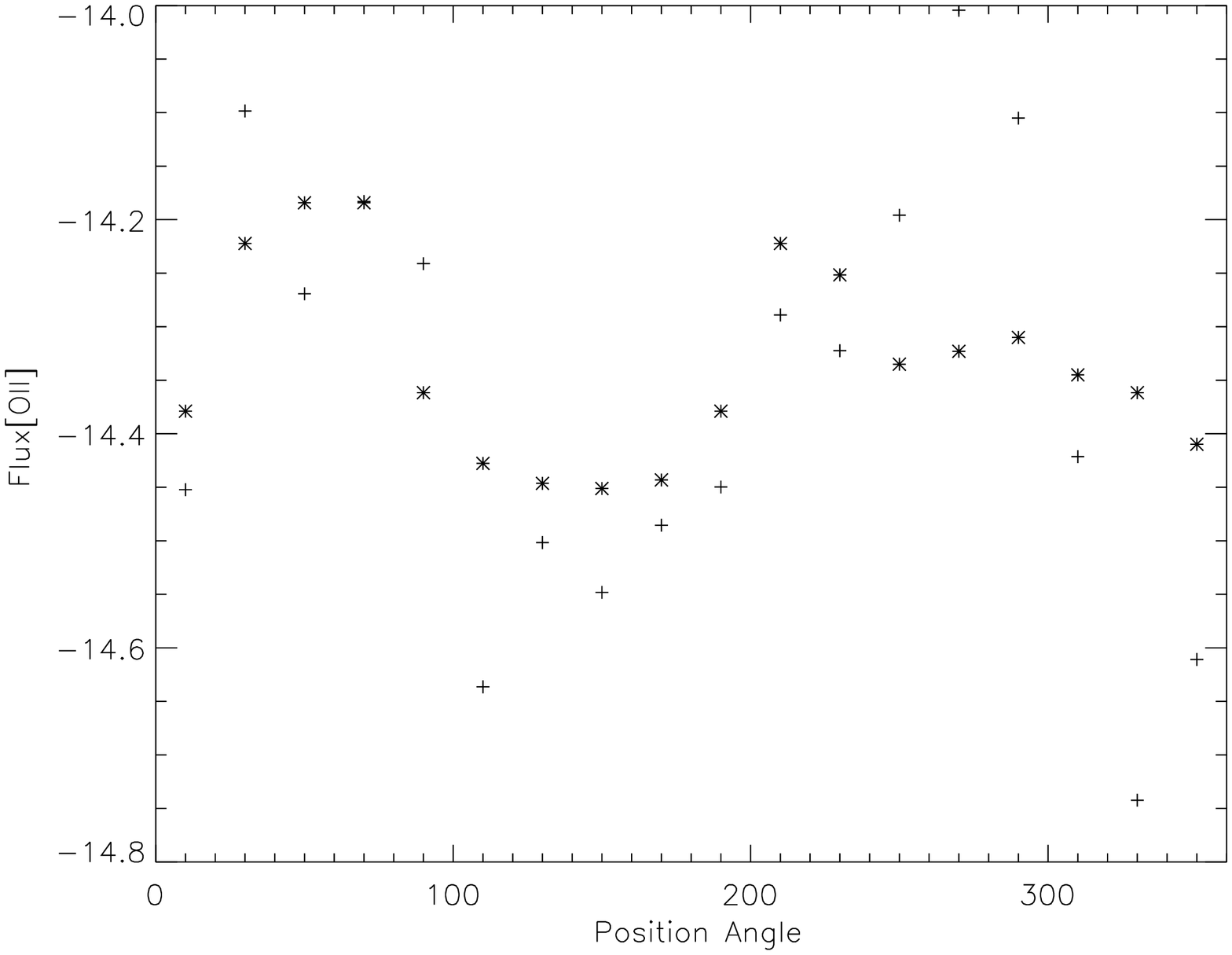}
\caption{{\it Top Panel}: Observed (crosses) and predicted (asterisks) [O~III] fluxes over the
range in position angle (PA) sampled at a radial distance of 30 pc. {\it Bottom Panel}: Observed and predicted [O~II]
fluxes for the same range in PA.}
\end{figure}

\clearpage

\begin{figure}
\includegraphics[scale=0.6,angle=90]{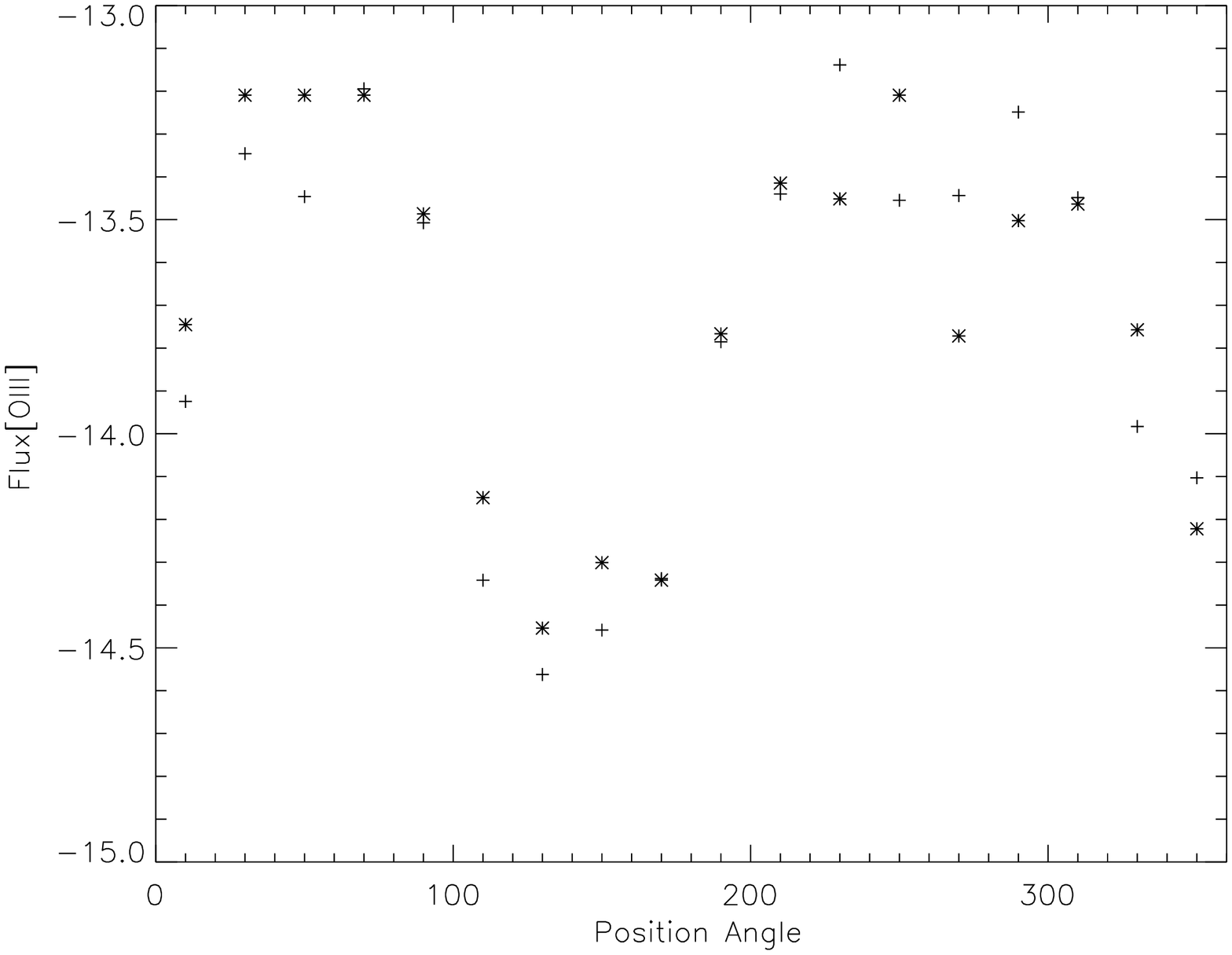}
\includegraphics[scale=0.6,angle=90]{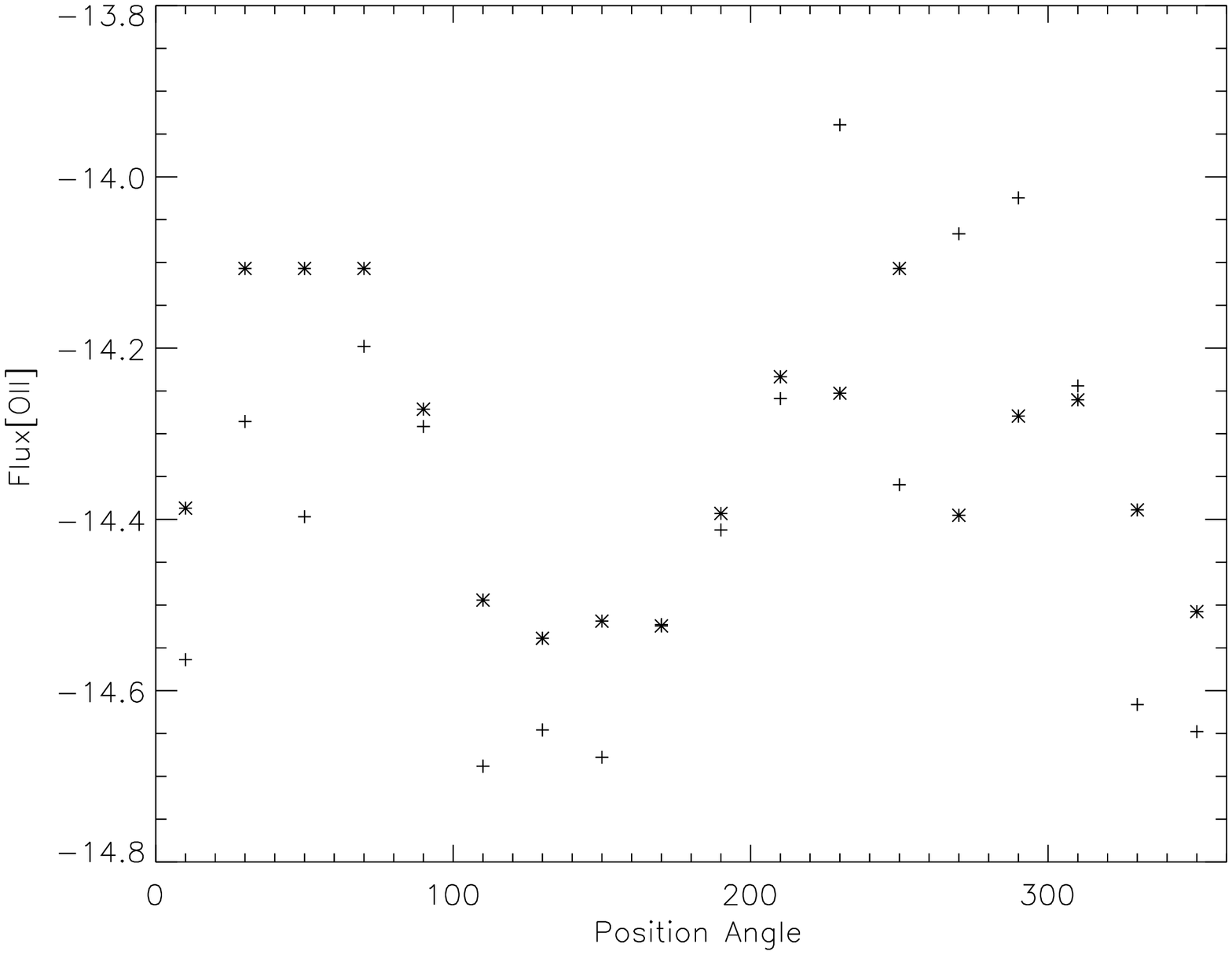}
\caption{{\it Top Panel}: Observed (crosses) and predicted (asterisks) [O~III] fluxes over the
range in position angle (PA) sampled at a radial distance of 38 pc. {\it Bottom Panel}: Observed and predicted [O~II]
fluxes for the same range in PA.}
\end{figure}

\clearpage

\begin{figure}
\includegraphics[scale=0.6,angle=90]{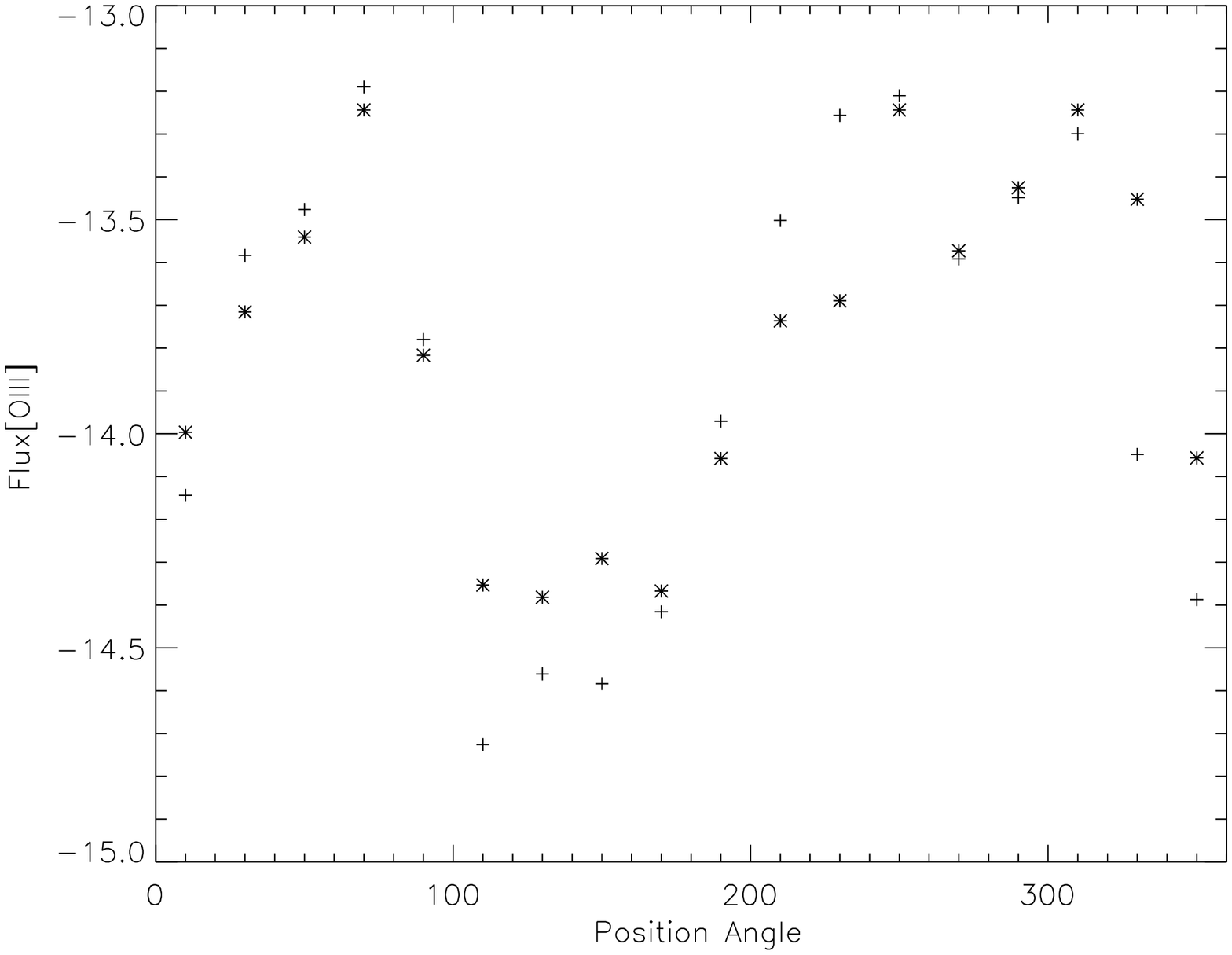}
\includegraphics[scale=0.6,angle=90]{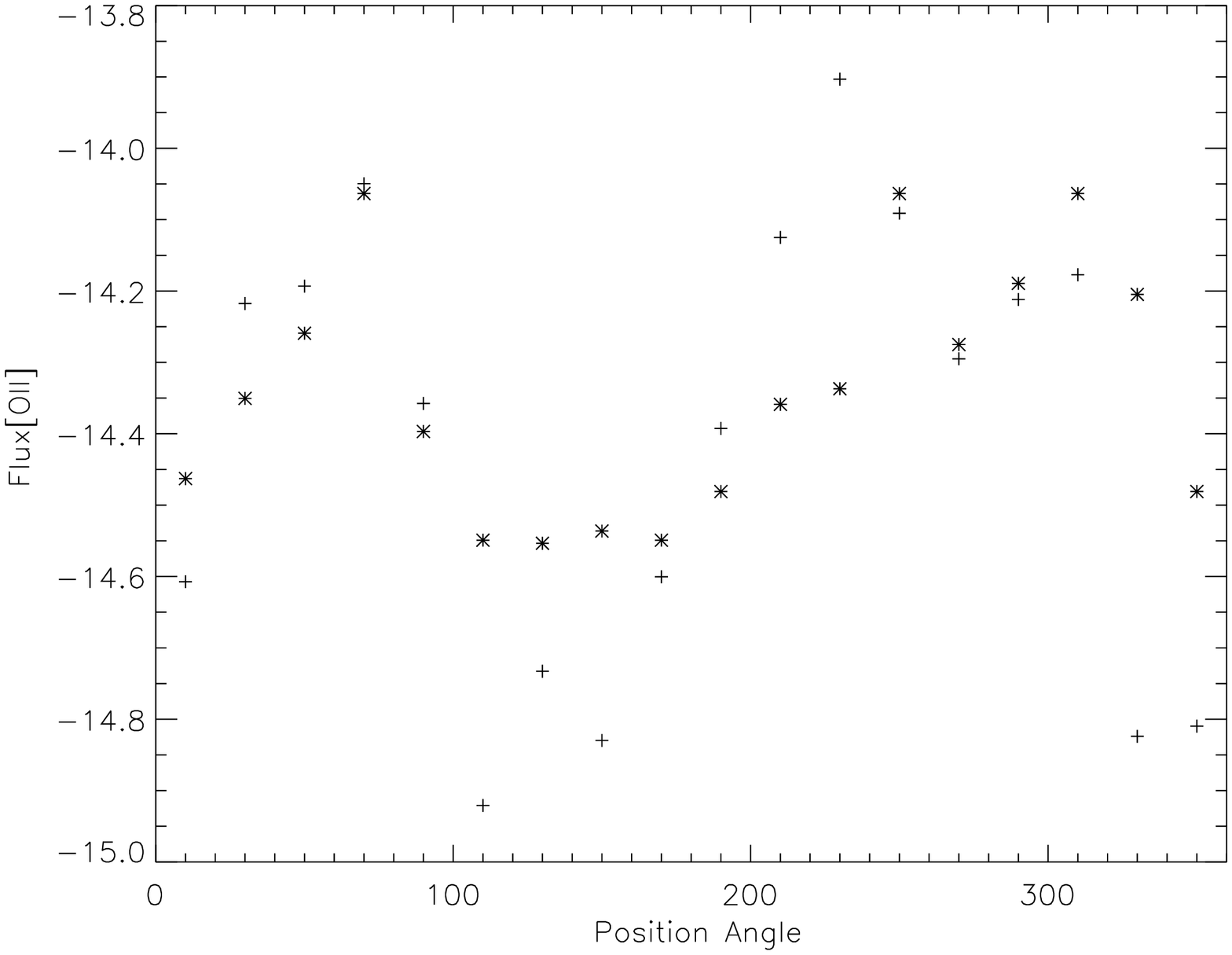}
\caption{{\it Top Panel}: Observed (crosses) and predicted (asterisks) [O~III] fluxes over the
range in position angle (PA) sampled at a radial distance of 47 pc. {\it Bottom Panel}: Observed and predicted [O~II]
fluxes for the same range in PA.}
\end{figure}

\clearpage

\begin{figure}
\includegraphics[scale=0.6,angle=90]{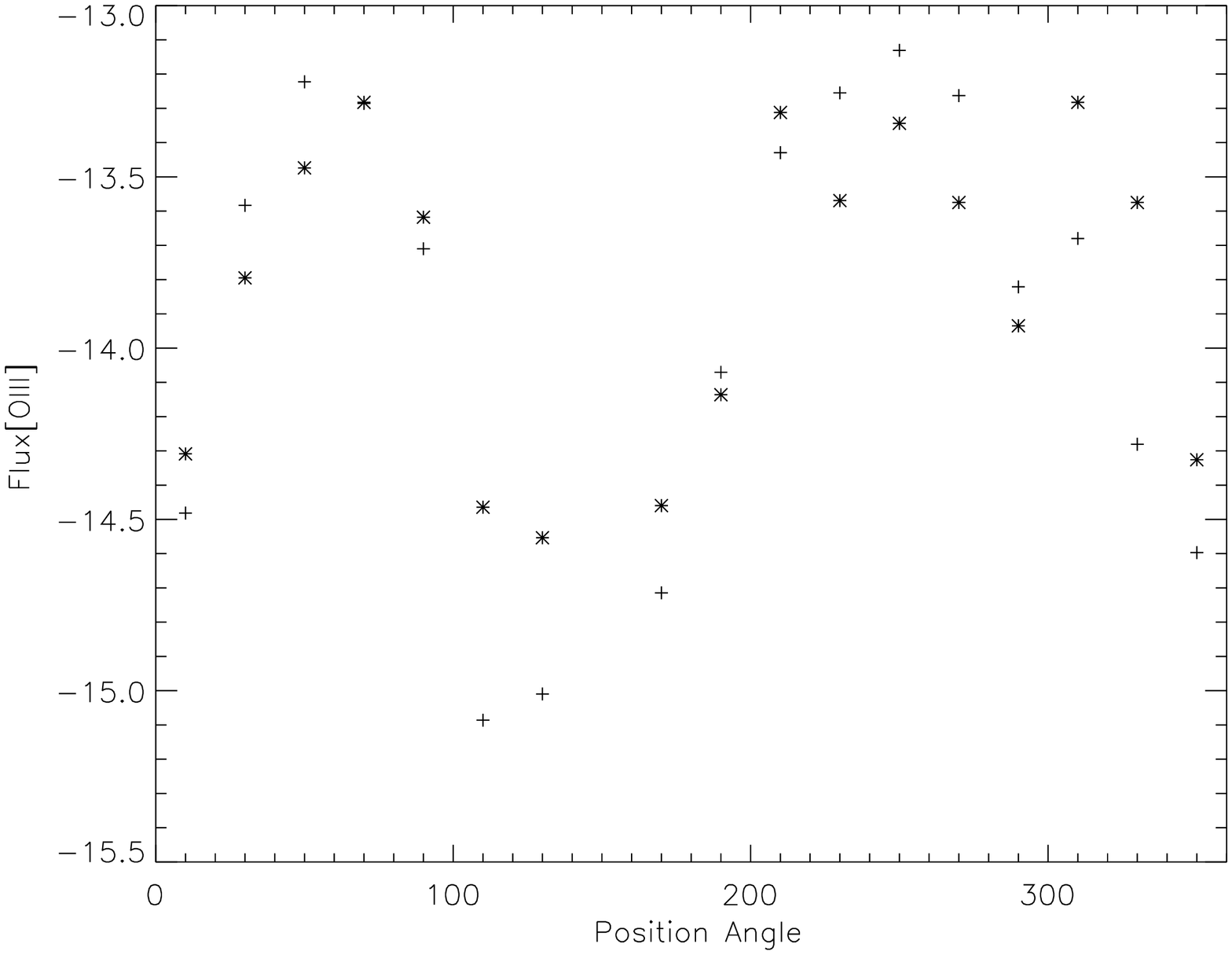}
\includegraphics[scale=0.6,angle=90]{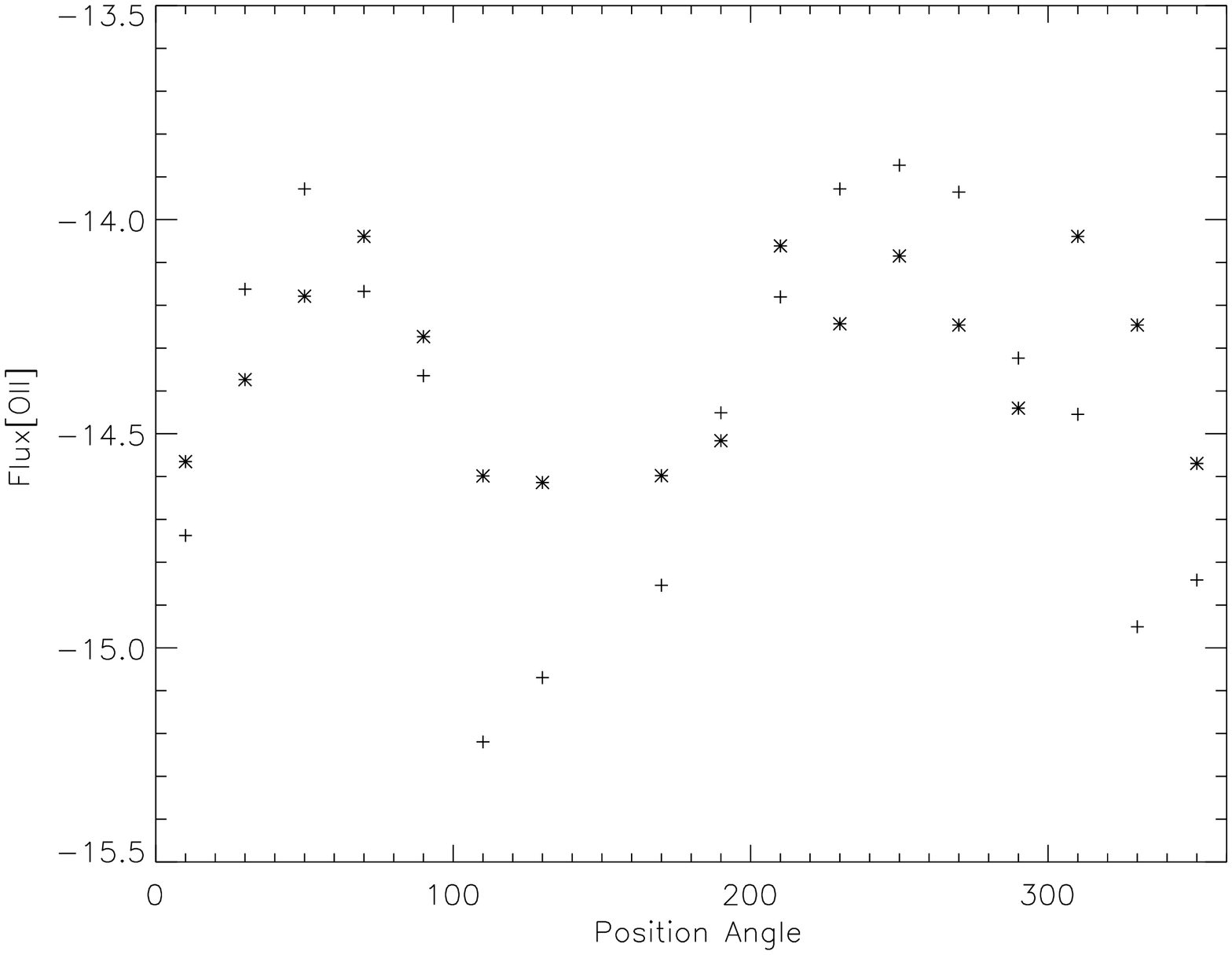}
\caption{{\it Top Panel}: Observed (crosses) and predicted (asterisks) [O~III] fluxes over the
range in position angle (PA) sampled at a radial distance of 56 pc. {\it Bottom Panel}: Observed and predicted [O~II]
fluxes for the same range in PA.}
\end{figure}

\clearpage

\begin{figure}
\includegraphics[scale=0.6,angle=90]{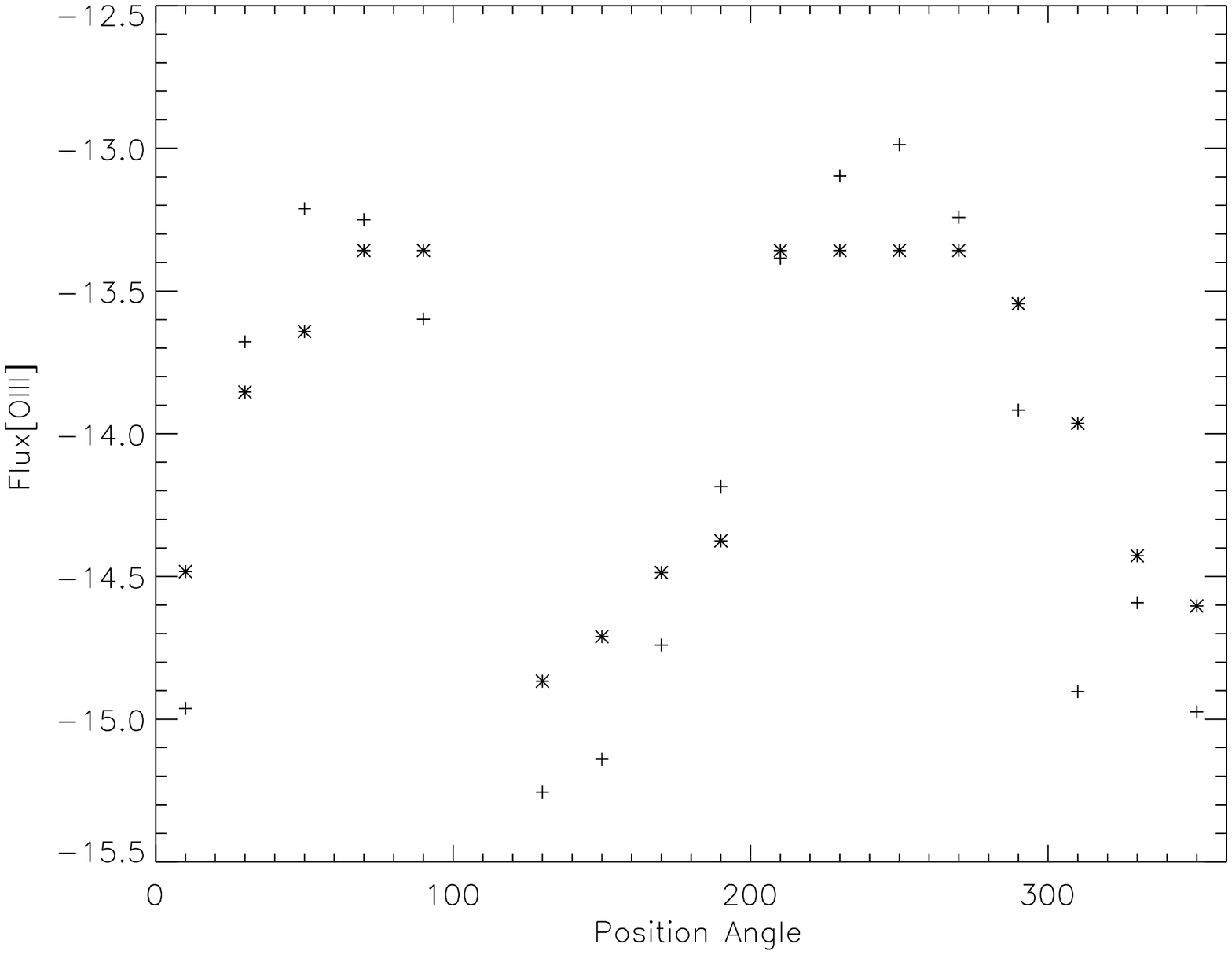}
\includegraphics[scale=0.6,angle=90]{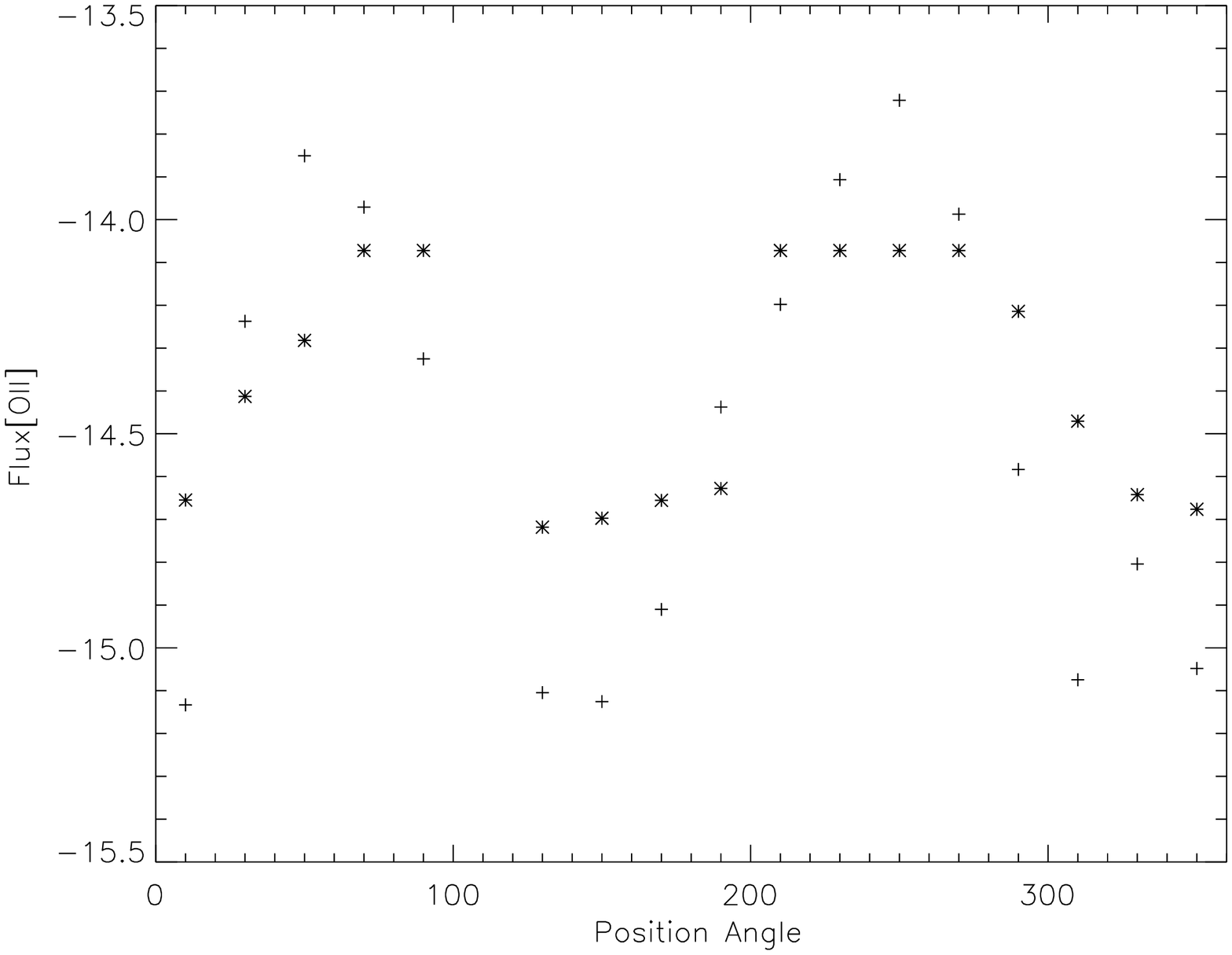}
\caption{{\it Top Panel}: Observed (crosses) and predicted (asterisks) [O~III] fluxes over the
range in position angle (PA) sampled at a radial distance of 64 pc. {\it Bottom Panel}: Observed and predicted [O~II]
fluxes for the same range in PA.}
\end{figure}

\clearpage

\begin{figure}
\includegraphics[scale=0.6,angle=90]{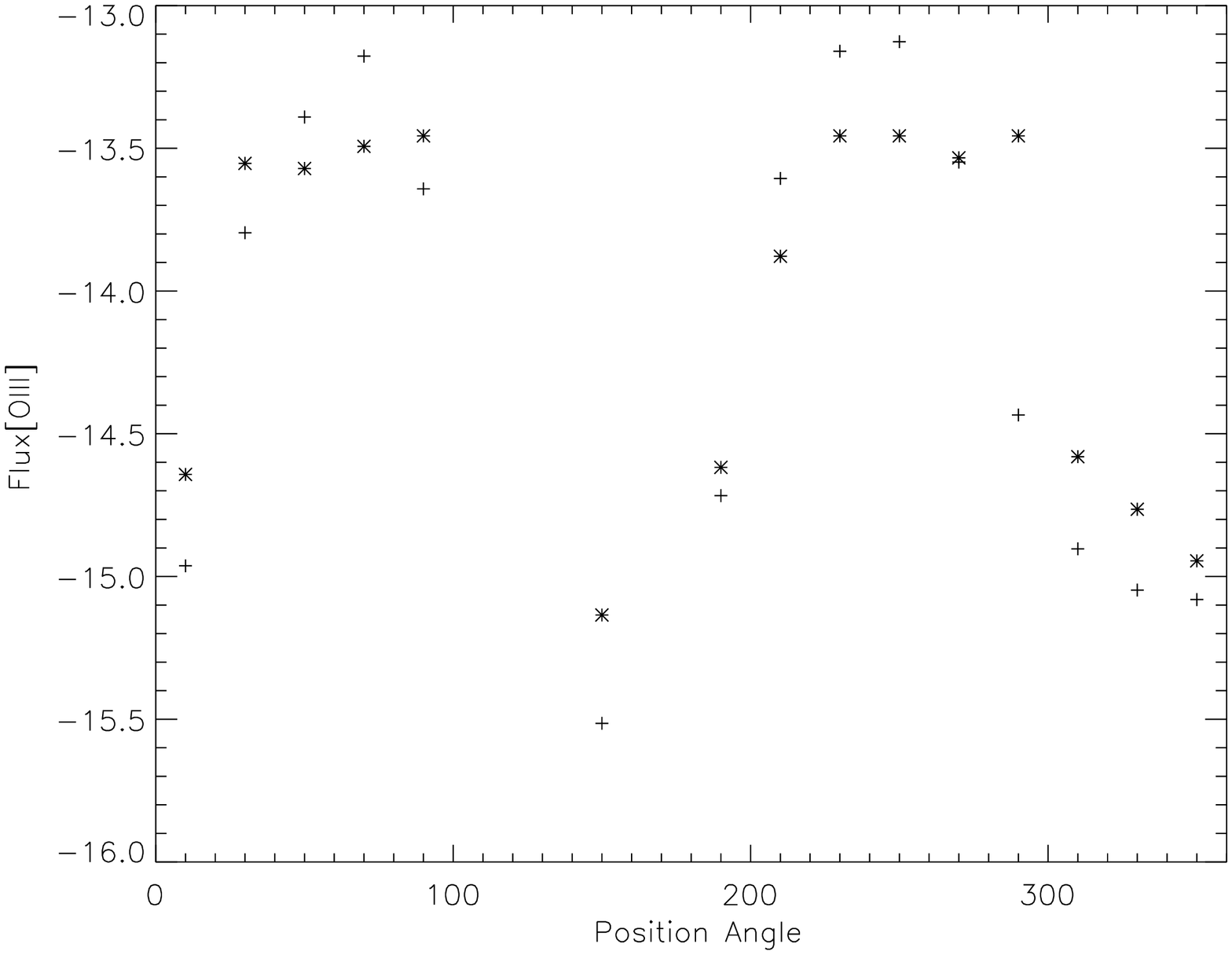}
\includegraphics[scale=0.6,angle=90]{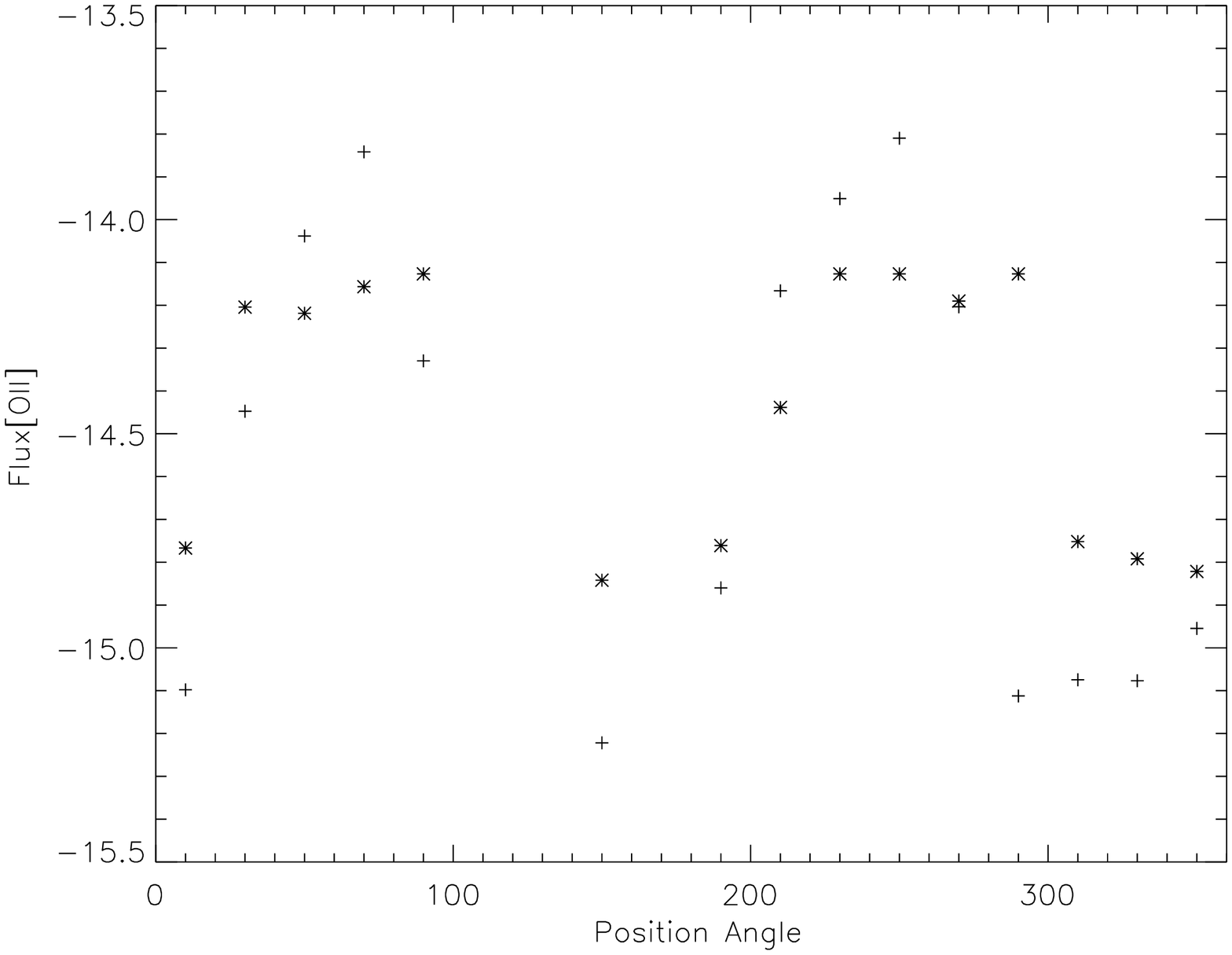}
\caption{{\it Top Panel}: Observed (crosses) and predicted (asterisks) [O~III] fluxes over the
range in position angle (PA) sampled at a radial distance of 73 pc. {\it Bottom Panel}: Observed and predicted [O~II]
fluxes for the same range in PA.}
\end{figure}

\clearpage

\begin{figure}
\includegraphics[scale=0.6,angle=90]{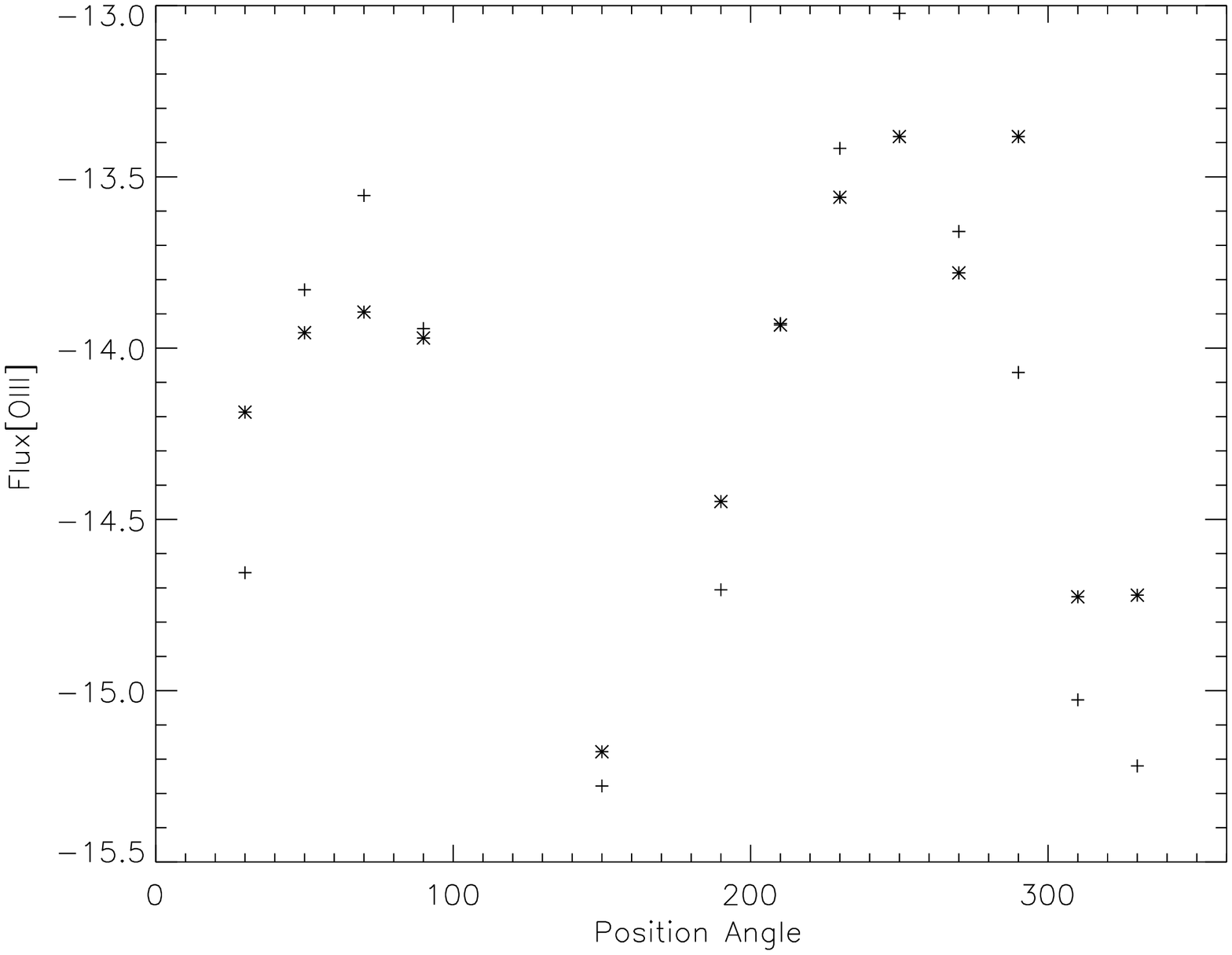}
\includegraphics[scale=0.6,angle=90]{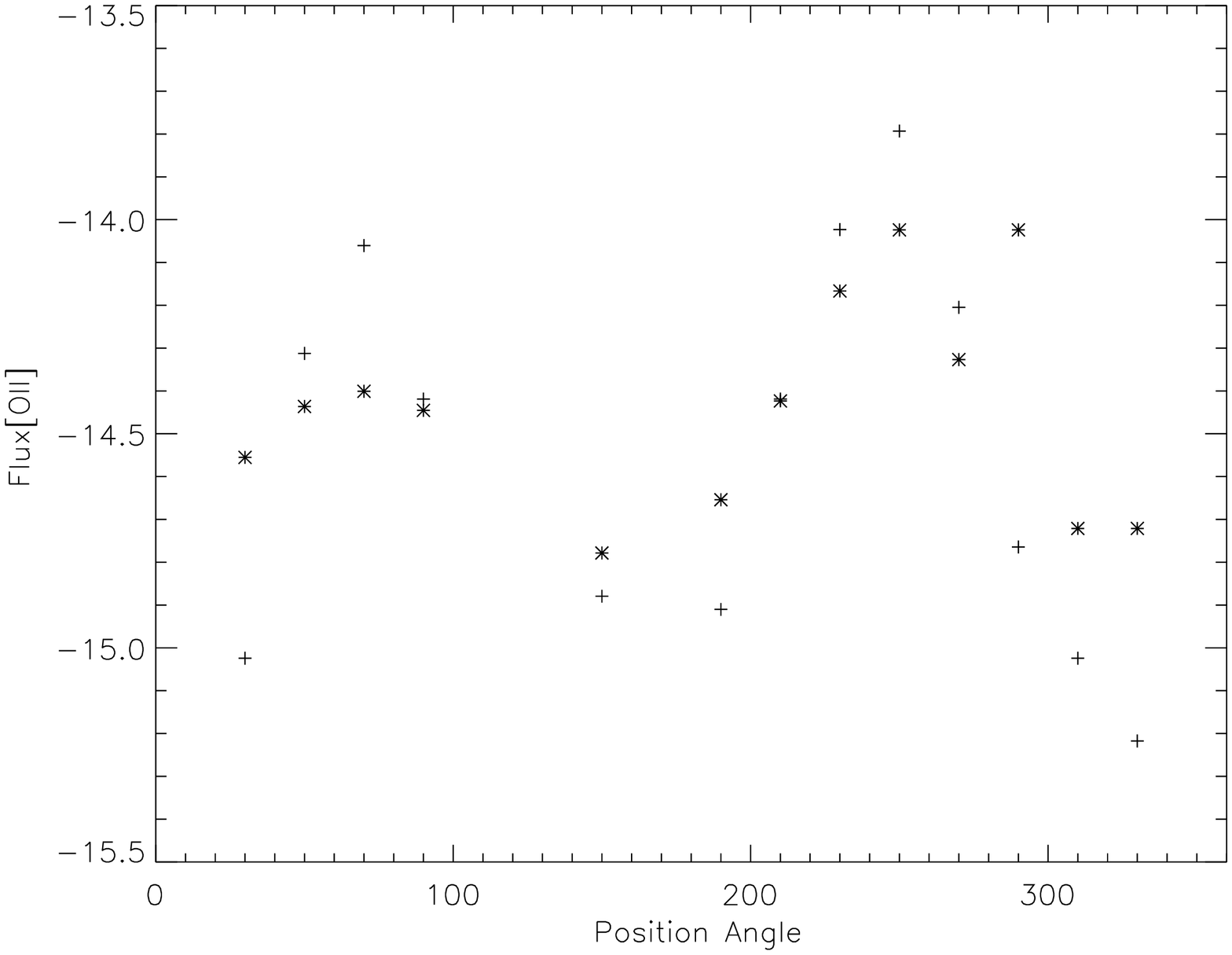}
\caption{{\it Top Panel}: Observed (crosses) and predicted (asterisks) [O~III] fluxes over the
range in position angle (PA) sampled at a radial distance of 81 pc. {\it Bottom Panel}: Observed and predicted [O~II]
fluxes for the same range in PA.}
\end{figure}

\clearpage

\begin{figure}
\includegraphics[scale=0.6,angle=90]{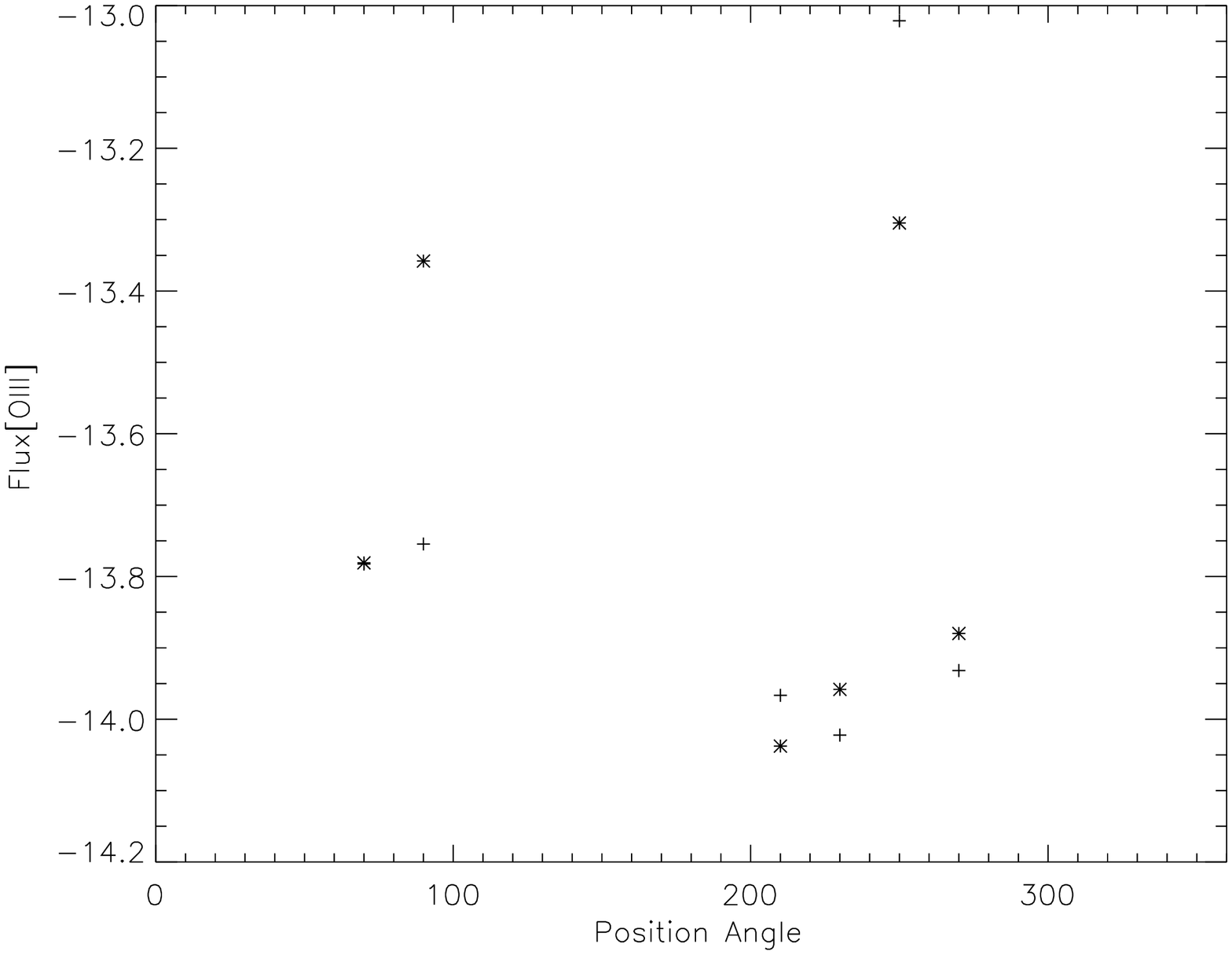}
\includegraphics[scale=0.6,angle=90]{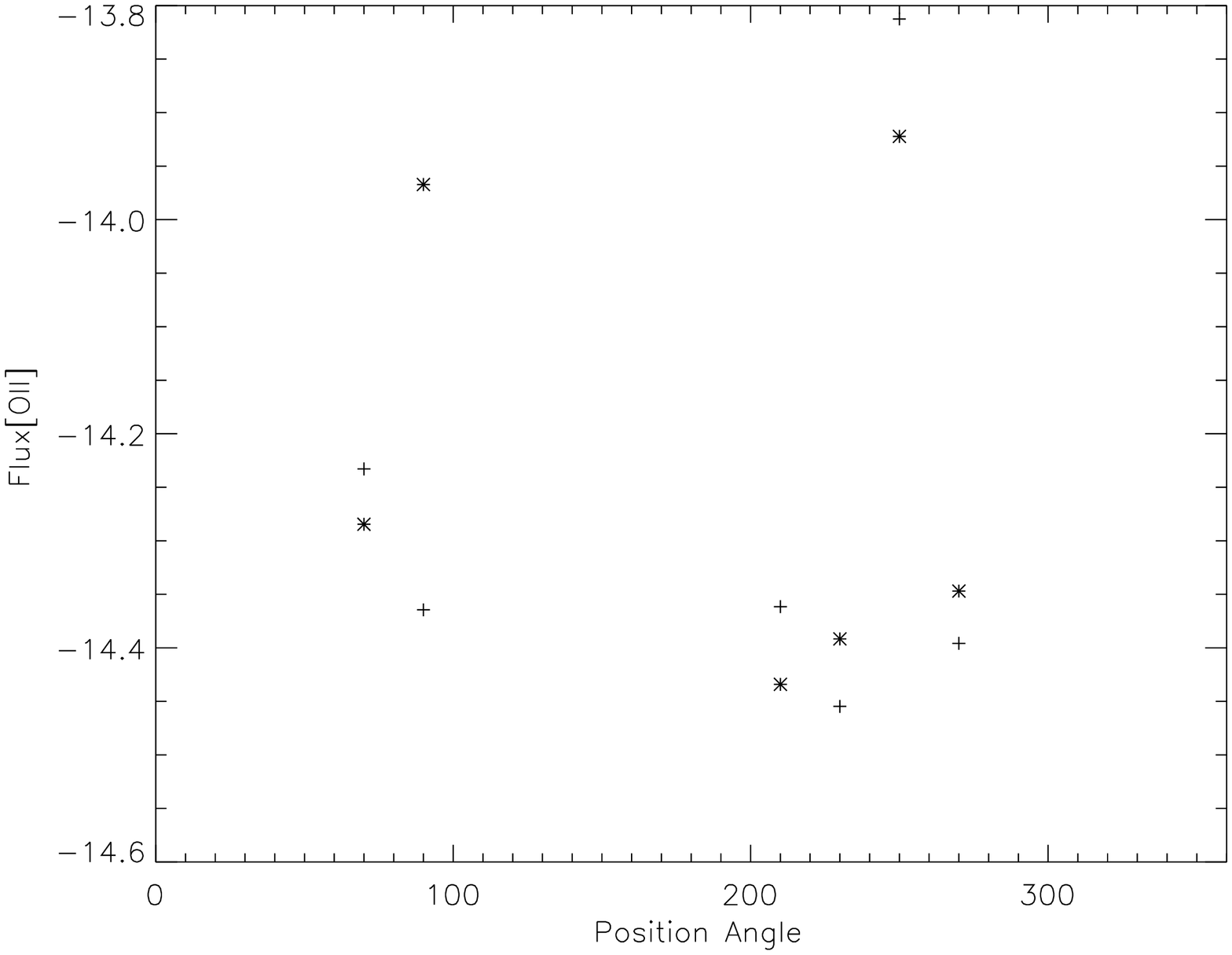}
\caption{{\it Top Panel}: Observed (crosses) and predicted (asterisks) [O~III] fluxes over the
range in position angle (PA) sampled at a radial distance of 90 pc. {\it Bottom Panel}: Observed and predicted [O~II]
fluxes for the same range in PA.}
\end{figure}

\clearpage

\begin{figure}
\includegraphics[scale=0.75,angle=0]{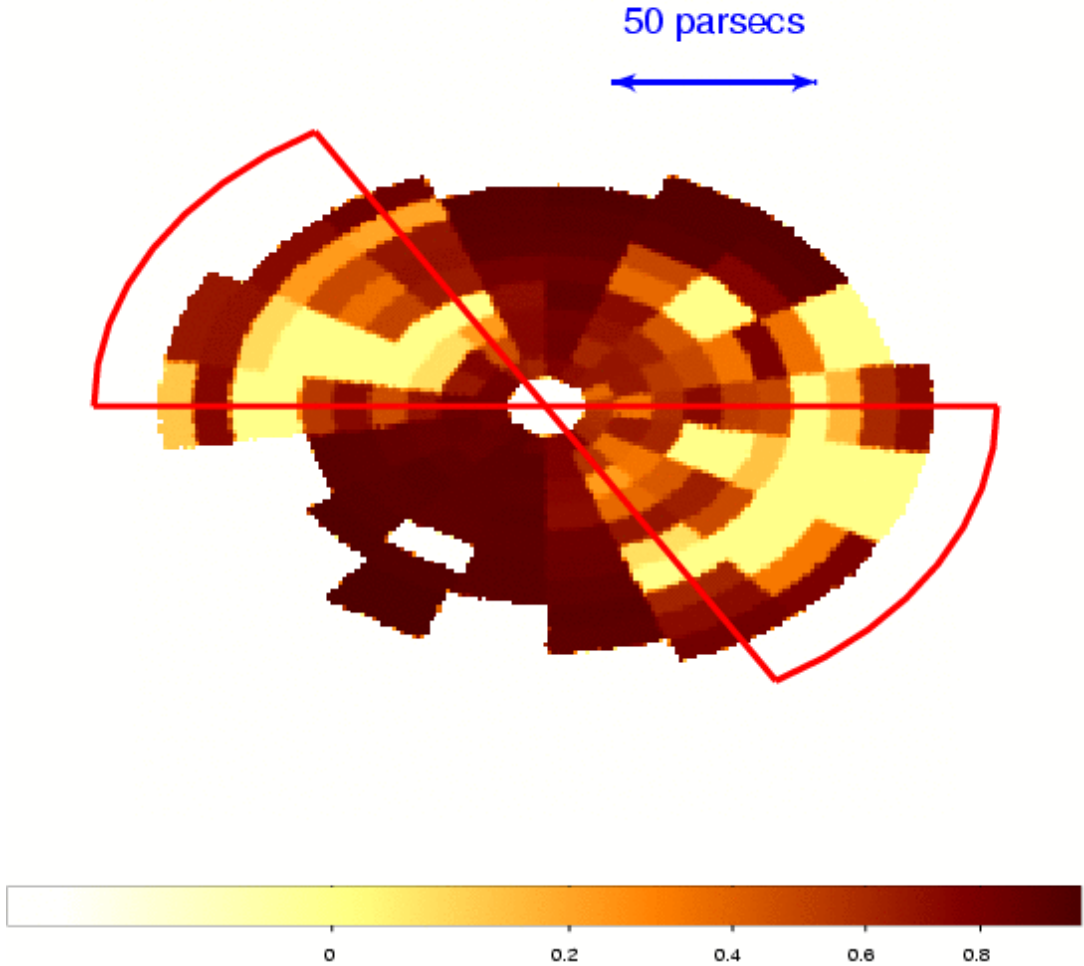}
\caption{Covering factor of LABS for each of the segments modeled. The wedges within the bi-cones with
the lowest covering factors for LABS are centered on PAs 70$^{\rm o}$ and 250$^{\rm o}$, on the left and right hand sides
of the nucleus, respectively. Note that, while the emission-line gas within the bi-cone is exposed to 
more of the unfiltered continuum than
the gas outside the bi-cone, there is significant radial variation in the covering factor
of LABS at some PAs.}
\end{figure}

\clearpage

\begin{figure}
\includegraphics[scale=0.75,angle=90]{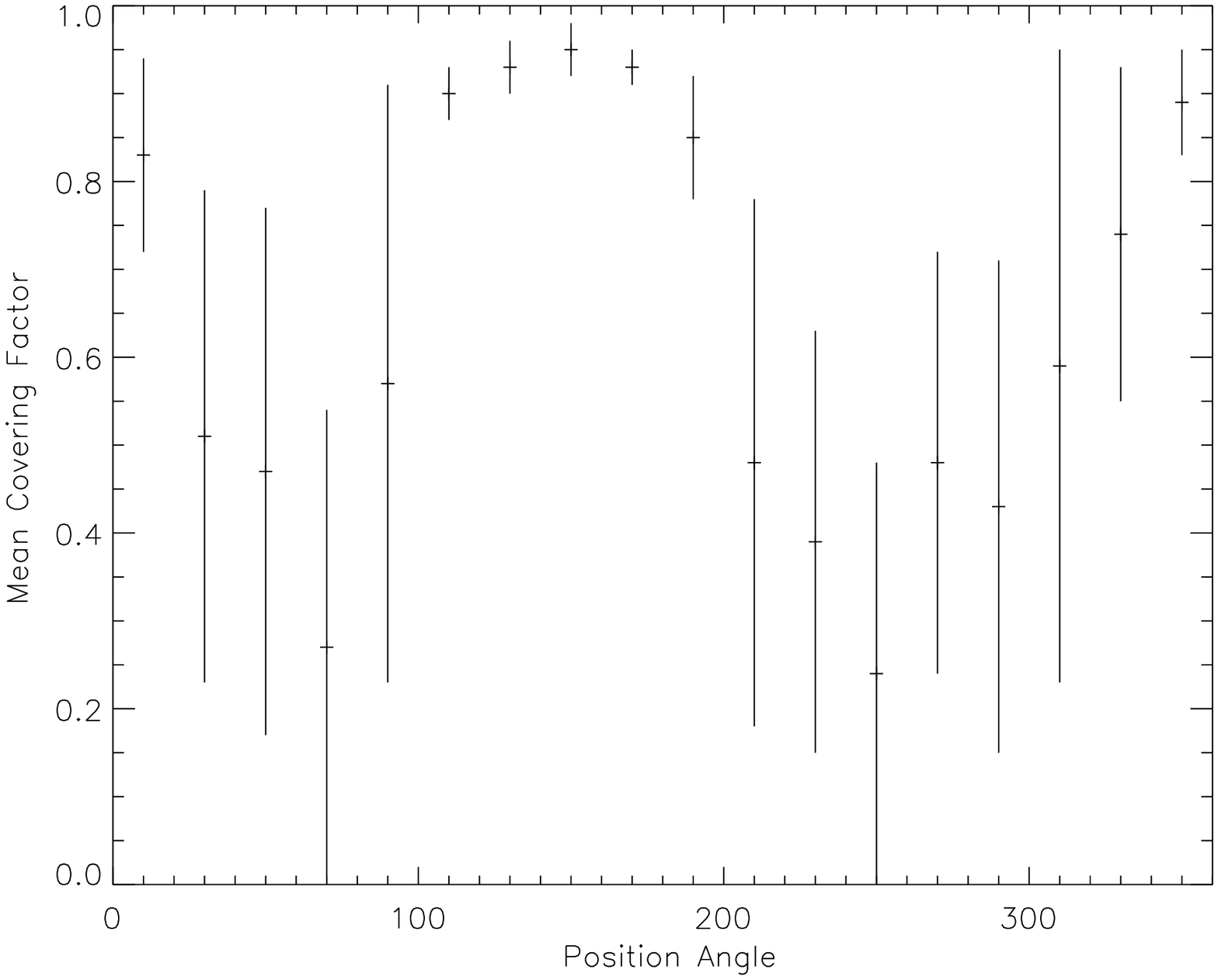}
\caption{Mean covering factors and standard deviations of LABS as a function of PA. Note that 
there are several regions with large
standard deviations, e.g. PA 310\deg~ (see text).}
\end{figure}


\begin{references}

\reference{and1969}Anderson, K.S., \& Kraft, R.P. 1969, \apj, 158, 859

\reference{ale1999}Alexander, T., Sturm, E., Lutz, D., Sternberg, A., Netzer, H., 
\& Genzel, R. 1999, \apj, 512, 204

\reference{ant1993}Antonucci, R.R.J. 1993, ARA\&A, 31, 473

\reference{ant1985}Antonucci, R.R.J., \& Miller, J.S. 1985, \apj, 297, 621

\reference{asp2005}Asplund, M., Grevesse, N., \& Sauval, A.J. 2005, in ASP
Conference Series 336, Cosmic Abundances as Records of Stellar Evolution and
Nucleosyntheses, ed. T.G. Barnes, III, \& F.N. Bash (San Francisco:ASP), 25

\reference{bar1977}Barr, P., White, N.E., Sanford, P.W., \& Ives, J.C. 1977, MNRAS, 181, 43P

\reference{bar1987}Barvainis, R. 1987, \apj, 320, 537

\reference{bok1978}Boksenberg, A., et al. 1978, Nature, 275, 404

\reference{cap1996}Capetti, A., Axon, D.J., Macchetto, F., Sparks, W.B., \& Boksenberg, A.
1996, \apj, 469, 554

\reference{cap1995}Capetti, A., Macchetto, F.D., Axon, D.J., Sparks, W.B., \& Boksenberg, A,
1995, \apj, 452, L87

\reference{cre2005}Crenshaw, D.M., \& Kraemer, S.B. 2005, \apj, 625, 680

\reference{cre2007}Crenshaw, D.M., \& Kraemer, S.B. 2007, \apj, 659, 250

\reference{cre1999}Crenshaw, D.M., Kraemer, S.B., Boggess, A., Maran, S.P., Mushotzky, R.F.,
\& Wu, C.-C. 1999, ApJ, 516, 750

\reference{cre2003}Crenshaw, D.M., Kraemer, S.B., \& George, I.M. 2003, ARA\&A, 41, 117

\reference{cre2000}Crenshaw, D.M., Kraemer, S.B., Hutchings, J.B., Bradley, L.D.
II, Gull, T.R., Kaiser, M.E., Nelson, C.H., Ruiz, J.R., \& Weistrop D. 2000, AJ,
120, 1731

\reference{das2005}Das, V., Crenshaw, D.M., Hutchings, J.B., Deo, R.P., 
Kraemer, S.B., Gull, T.R., Kaiser, M.E., Nelson, C.H., \& Weistrop, D.
2005, AJ, 130, 945

\reference{dop1995}Dopita, M.A., \& Sutherland, R.S. 1995, \apj, 455, 468

\reference{eva1993}Evans, I. N., Tsvetanov, Z., Kriss, G. A.,
Ford, H. C., Caganoff, S., \& Koratkar, A. P. 1993, ApJ, 417, 82

\reference{fel1999}Feldmeier, J.J., Brandt, W.N., Elvis, M., Fabian, A.C.,
Iwasawa, K., \& Mathur, S. 1999, \apj, 510, 167


\reference{fer1998}Ferland, G.J., Korista, K.T., Verner, D.A., Ferguson, J.W., Kingdon, J.B.,
\& Verner, E.M. 1998, PASP, 110, 749

\reference{fer1983}Ferland, G.J., \& Netzer, H. 1983, \apj, 264, 105

\reference{hol1980}Holt, S.S., Mushotzky, R.F., Becker, R.H., Boldt, E.A., Serlemitsos, P.J.,
Szymkowiak, A.E., \& White, N.E. 1980, \apj, 241, L13

\reference{hut1998}Hutchings, J.B., et al. 1998, \apj, 492, L115

\reference{geo1998}George, I.M., Turner, T.J., Netzer, H., Nandra, K., Mushotzky, R.F.,
\& Yaqoob, T. 1998, ApJS, 114, 73

\reference{gro2004}Groves, B.A., Dopita, M.A., \& Sutherland, R.S. 2004, ApJS, 153, 9

\reference{kai2000}Kaiser, M.E., Bradley, L.D. II, Hutchings, J.B., Crenshaw,
D.M., Gull, T.R., Kraemer, S.B., Nelson, C.H., Ruiz, J., \& Weistrop, D. 2000,
\apj, 528, 260

\reference{kha1971}Khachikian, E. Ye., \& Weedman, D.W. 1971, Astrofizika, 7, 389

\reference{kin2002}Kinkhabwala, A., Sako, M., Behar, E., Kahn, S.M., Paerels,
F., Brinkman, A.C>, Kaastra, J.S., Gu, M.F., \& Liedahl, D.A. 2002, \apj, 575,
732

\reference{kon1994}Konigl, A., \& Kartje, J.F. 1994, \apj, 434, 446

\reference{kri1992}Kriss, G.A., et al. 1992, ApJ, 394, L37

\reference{kri1995}Kriss, G.A., Davidsen, A.F., Zheng, W., Kruk, J.W., \& Espey, B.R.
1995, \apj, 454, L7

\reference{kra2000}Kraemer, S.B., \& Crenshaw, D.M. 2000a, \apj, 532, 256

\reference{kra2000}Kraemer, S.B., \& Crenshaw, D.M. 2000b, \apj, 544, 763


\reference{kra2000}Kraemer, S.B., Crenshaw, D.M., Hutchings, J.B., Gull, T.R., 
Kaiser, M.E., Nelson, C.H., \& Weistrop, D. 2000, \apj, 531, 278 (K2000)

\reference{kra2001}Kraemer, S.B., Crenshaw, D.M., Hutchings, J.B., 
Danks, A.C., Gull, T.R., Kaiser, M.E., Nelson, C.H., \& Weistrop, D. 2001, 
\apj, 551, 671 


\reference{kra2005}Kraemer, S.B., et al. 2005, \apj, 633, 693

\reference{kra2006}Kraemer, S.B., et al. 2006, ApJS, 167, 161

\reference{kro1988}Krolik, J.H., \&  Begelman, M.C. 1988, \apj, 329, 702


\reference{nel2000}Nelson, C.H., Weistrop, D., Hutchings, J.B., Crenshaw, D.M.,
Gull, T.R., Kaiser, M.E., Kraemer, S.B., \& Lindler, D. 2000, \apj, 531, 275

\reference{oke1968}Oke, J.B., \& Sargent, W.L.W. 1968, \apj, 151, 807

\reference{ost1989}Osterbrock, D.E., 1989, Astrophysics of Gaseous Nebulae
and Active Galactic Nuclei (Mill Valley: University Science Books)

\reference{pet2004}Peterson, B.M., et al. 2004, \apj, 613, 682

\reference{pog1989}Pogge, R. W. 1989, \apj, 345, 730


\reference{sak2000}Sako, M., Kahn, S.M., Paerels, F., \& Liedahl, D.A. 2000,
\apj, 543, L115 

\reference{sch2003}Schmitt, H.R., Donley, J.L, Antonucci, R.R.J., Hutchings, J.B.,
\& Kinney, A.L. 2003, ApJS, 148, 327


\reference{sch1996}Schmitt, H.R., \& Kinney, A.L. 1996, \apj, 463, 498

\reference{sch2002}Schmitt, H.R., Kinney, A.L, Hutchings, J.B., Ulvestad, J.S.,
\& Antonucci, R.R.J. 2002, in ASP Conf. Ser. 255,  Mass Outflow in Active 
Galactic Nuclei: New
Perspectives, ed. D.M. Crenshaw, S.B. Kraemer, \& I.M. George (San Francisco:
ASP), 215

\reference{sch1993}Schulz, H., \& Komossa, S. 1993, A\&A, 278, 29

\reference{wey1982}Weymann, R.J., Scott, J.S., Schiano, A.V.R., \& Christiansen, W.A.
1982, \apj, 262, 497


\reference{yaq1989}Yaqoob, T., Warwick, R.S., \& Pounds, K.A. 1989, MNRAS, 236, 153

\end{references}
\end{document}